\documentclass[12pt]{article}

\usepackage[margin=1in]{geometry}

\usepackage{setspace}

\usepackage{graphicx}

\usepackage{enumitem}

\usepackage{microtype}
\usepackage{subfigure}
\usepackage{booktabs}
\usepackage{comment}
\usepackage{wrapfig}
\usepackage{arydshln}
\usepackage{enumitem}
\usepackage{multirow}
\usepackage{url}
\usepackage{natbib}
\usepackage{authblk}

\RequirePackage[colorlinks,citecolor=blue,linkcolor=blue,urlcolor=blue,pagebackref]{hyperref}

\usepackage{amsmath}
\usepackage{amssymb}
\usepackage{mathtools}
\usepackage{amsfonts}
\usepackage{bm}
\usepackage{bbm}

\usepackage{algorithm}
\usepackage{algorithmic}

\def\eqref#1{equation~\ref{#1}}
\def\ceil#1{\lceil #1 \rceil}

\def\1{\bm{1}}

\def\rmd{{\mathrm{d}}}

\def\bmmu{{\bm{\mu}}}
\def\bmnu{{\bm{\nu}}}

\def\bmm{{\bm{m}}}

\def\bmP{{\bm{P}}}

\def\bmmu{{\bm{\mu}}}
\def\bmnu{{\bm{\nu}}}

\DeclareMathAlphabet{\mathsfit}{\encodingdefault}{\sfdefault}{m}{sl}
\SetMathAlphabet{\mathsfit}{bold}{\encodingdefault}{\sfdefault}{bx}{n}

\def\calA{{\mathcal{A}}}
\def\calB{{\mathcal{B}}}

\def\calE{{\mathcal{E}}}
\def\calF{{\mathcal{F}}}

\def\calH{{\mathcal{H}}}

\def\calM{{\mathcal{M}}}
\def\calN{{\mathcal{N}}}

\def\calP{{\mathcal{P}}}

\def\calY{{\mathcal{Y}}}

\def\bbE{{\mathbb{E}}}

\def\bbP{{\mathbb{P}}}

\def\bbR{{\mathbb{R}}}

\DeclareMathOperator*{\argmax}{arg\,max}

\newcommand{\p}[1]{\left(#1\right)}
\newcommand{\sqb}[1]{\left[#1\right]}
\newcommand{\cb}[1]{\left\{#1\right\}}

\newcommand{\bigp}[1]{\big(#1\big)}
\newcommand{\bigsqb}[1]{\big[#1\big]}
\newcommand{\bigcb}[1]{\big\{#1\big\}}

\newcommand{\Bigp}[1]{\Big(#1\Big)}
\newcommand{\Bigsqb}[1]{\Big[#1\Big]}
\newcommand{\Bigcb}[1]{\Big\{#1\Big\}}

\newcommand{\Biggp}[1]{\Bigg(#1\Bigg)}

\newcommand{\Biggcb}[1]{\Bigg\{#1\Bigg\}}

\newcommand{\abs}[1]{\left|#1\right|}
\newcommand{\bigabs}[1]{\big|#1\big|}

\usepackage{amsthm}

\theoremstyle{plain}

\newtheorem{theorem}{Theorem}[section]
\newtheorem{lemma}[theorem]{Lemma}

\newtheorem{corollary}[theorem]{Corollary}
\newtheorem{proposition}[theorem]{Proposition}

\newtheorem{definition}{Definition}[section]
\newtheorem{assumption}{Assumption}[section]
\newtheorem*{example}{Example}
\newtheorem*{remark}{Remark}

\usepackage[textsize=tiny]{todonotes}
\usepackage{multirow}
\usepackage{wrapfig}
\usepackage{subfigure}

\renewcommand{\eqref}[1]{(\ref{#1})}

\usepackage{xcolor}
\newcount\Comments  % 0 suppresses notes to selves in text
\newcommand{\kibitz}[2]{\ifnum\Comments=1\textcolor{#1}{#2}\fi}

\usepackage[capitalize,noabbrev]{cleveref}

\allowdisplaybreaks

\title{Minimax and Bayes Optimal Adaptive Experimental Design for Treatment Choice}

\author{Masahiro Kato\thanks{Email: \texttt{mkato-csecon@g.ecc.u-tokyo.ac.jp}}$\,$}

\affil{The University of Tokyo}

\date{\today}
\begin{document}

\maketitle 

\begin{abstract}
We consider an adaptive experiment for treatment choice and design a minimax and Bayes optimal adaptive experiment with respect to regret. Given binary treatments, the experimenter's goal is to choose the treatment with the highest expected outcome through an adaptive experiment, in order to maximize welfare. We consider adaptive experiments that consist of two phases, the treatment allocation phase and the treatment choice phase. The experiment starts with the treatment allocation phase, where the experimenter allocates treatments to experimental subjects to gather observations. During this phase, the experimenter can adaptively update the allocation probabilities using the observations obtained in the experiment. After the allocation phase, the experimenter proceeds to the treatment choice phase, where one of the treatments is selected as the best. For this adaptive experimental procedure, we propose an adaptive experiment that splits the treatment allocation phase into two stages, where we first estimate the standard deviations and then allocate each treatment proportionally to its standard deviation. We show that this experiment, often referred to as Neyman allocation, is minimax and Bayes optimal in the sense that its regret upper bounds exactly match the lower bounds that we derive. To show this optimality, we derive minimax and Bayes lower bounds for the regret using change-of-measure arguments. Then, we evaluate the corresponding upper bounds using the central limit theorem and large deviation bounds.
\end{abstract}

\section{Introduction}
Statistical decision making for choosing a better treatment is a core topic in causal inference \citep{Manski2004statisticaltreatment}. This study investigates adaptive experimental design for treatment choice given binary treatments. In an adaptive experiment, we can adaptively update the treatment allocation probabilities during an experiment to gather samples and efficiently choose a better treatment at the end of the experiment. This problem has been investigated under various names, such as (fixed-budget) \emph{best-arm identification} \citep[BAI,][]{Audibert2010bestarm}, ordinal optimization \citep{Chen2000simulationbudget,Glynn2004largedeviations}, and adaptive experimental design for policy choice \citep{Kasy2021adaptivetreatment}. 

This study formulates an adaptive experimental procedure with two phases \citep{Kaufmann2016complexity}, the allocation phase and the treatment choice phase. Given a total of $T$ rounds, an experimenter allocates a treatment at each round, utilizing the data obtained up to that round. At the end of the experiment, the experimenter identifies the treatment with the highest expected outcome and recommends it for the future population. We measure the performance of the adaptive experiment using the regret, also called the simple regret, out-of-sample regret, or policy regret, which is defined as the difference between the expected outcomes under the best and the chosen treatments. 

For this setup, we design an adaptive experiment under which the regret is minimax optimal and Bayes optimal. In Section~\ref{sec:setup}, we formulate the problem. Then, in Section~\ref{sec:tsna}, we propose our adaptive experiment, called the two-stage Neyman allocation (TSNA) experiment. The TSNA experiment splits the allocation phase into two stages, where we uniformly randomly allocate each treatment in the first phase, and then allocate treatments proportional to the estimated standard deviations. To show the minimax and Bayes optimality of this experiment, we develop lower bounds in Section~\ref{sec:lowerbound}. Lastly, we investigate the worst-case and average upper bounds of the TSNA experiment and confirm that the minimax and Bayes regret upper bounds align with our derived lower bounds, which implies minimax and Bayes optimality.

\subsection{Setup}
\label{sec:setup}
We consider binary treatments, $1$ and $0$, where treatment $1$ is often called the treatment, and treatment $0$ is often called the control. For each treatment $d \in \{1,0\}$, let $Y_d \in \calY$ be a potential outcome, where $\calY \subseteq \bbR$ denotes the outcome space. Suppose that each potential outcome $Y_d$ follows a (marginal) distribution $P_{d, \mu_d}$ parameterized by $\mu_d \in \calM$, where $\calM \subset \bbR$ is a parameter space, which is convex and an open subset of $\bbR$. Given the parameter vector $\bmmu \coloneqq (\mu_1, \mu_0) \in \calM^2$, let $\bmP_{\bmmu} \coloneqq (P_{1, \mu_1}, P_{0, \mu_0})$ be a set of parametric distributions. The parameter $\mu_d$ is the mean of $Y_d$; that is, $\bbE_{\bmmu}[Y_d] = \mu_d$ holds, where $\bbE_{\bmmu}[\cdot]$ is the expectation under $\bmP_{\bmmu}$. Under a distribution $\bmP_{\bmmu}$, we consider the following adaptive experiment.

\paragraph{Adaptive experiment.} 
An experimenter aims to recommend a treatment with the highest expected outcome, denoted as
\[
    d^*_{\bmmu} = \arg\max_{d \in \{1, 0\}} \mu_d,
\]
through an adaptive experiment in which data are sampled from $\bmP_\bmmu$. 
Let $T$ denote the total sample size, also referred to as the budget. We consider an experimenter who follows the following adaptive experimental procedure consisting of two phases:
\begin{enumerate}
    \item \textbf{Treatment allocation phase}: For each $t \in [T] \coloneqq \{1, 2, \dots, T\}$:
    \begin{itemize}
        \item the experimenter allocates treatment $D_t \in \{1, 0\}$ to an experimental subject, based on the past observations $\{(D_s, Y_s)\}_{s=1}^{t-1}$,
        \item the experimenter observes the corresponding outcome $Y_t$, where $ Y_t \coloneqq \sum_{d \in \{1, 0\}} \mathbbm{1}[D_t = d]Y_{d, t}$, and $(Y_{d, t})_{d\in\{1, 0\}}$ follows the distribution $\bmP_{\bmmu}$.
    \end{itemize}
    \item \textbf{Treatment choice phase}: At the end of the experiment (after observing $Y_T$), based on the observed outcomes $\p{\bigp{D_t, Y_t}}_{t=1}^{T}$, the experimenter chooses treatment $\widehat{d}_T \in \{1, 0\}$ as an estimate of the best treatment $d^*_{\bmmu}$.
\end{enumerate}

Our task is to design an adaptive experiment $\delta$ that determines how the experimenter acts in the above procedure. That is, an adaptive experiment $\delta$ is a pair 
\[\p{\p{D_t^{\delta}}_{t \in [T]}, \widehat{d}_T^{\delta}},\]
where $\p{D_t^{\delta}}_{t \in [T]}$ are indicators for the allocated treatments in the treatment allocation phase, and $\widehat{d}_T^{\delta}$ is the estimator of the best treatment $d^*_{\bmmu}$ in the treatment choice phase. For simplicity, we omit the subscript $\delta$ when the dependence is clear from the context. 

\paragraph{Regret.} 
The performance of an adaptive experiment $\delta$ is measured by the regret, defined as:
\begin{align*}
    \text{Regret}^\delta_{\bmmu} \coloneqq \bbE_{\bmmu}\left[ Y_{d^*_{\bmmu}} - Y_{\widehat{d}_T^{\delta}} \right] = \bbE_{\bmmu}\sqb{\mu_{d^*_{\bmmu}} - \mu_{\widehat{d}_T^{\delta}}},
\end{align*}
where the expectation is taken over both $Y_d$ and $\widehat{d}_T^{\delta}$. 
This regret is referred to as the expected simple regret in the literature on BAI, where $Y_{d^*_{\bmmu}} - Y_{\widehat{d}_T^{\delta}}$ is called the simple regret. This regret is also referred to as the out-of-sample regret or the policy regret \citep{Kasy2021adaptivetreatment}. For simplicity, we refer to the expected simple regret as the simple regret in this study. An adaptive experiment $\delta$ with a smaller regret is considered a better experiment.

We focus on the worst-case regret and the regret averaged for a given prior $H(\bmmu)$ for the parameters $\bmmu$:
\begin{itemize}
    \item \textbf{Worst-case regret:} $\sup_{\bmmu \in \calM^2} \text{Regret}^\delta_{\bmmu}$. 
    \item \textbf{Average regret:} $\int_{\bmmu \in \calM^2} \text{Regret}^\delta_{\bmmu}\rmd H(\bmmu)$. 
\end{itemize}
We develop lower bounds for these two regrets, and experiments whose regrets match the lower bounds are called minimax and Bayes optimal.

\paragraph{Notation and assumptions.} Let $\bbP_{\bmmu}$ denote the probability law under $\bmP_{\bmmu}$, and let $\bbE_{\bmmu}$ denote the corresponding expectation operator. For each $d\in \{1, 0\}$, let $P_{d, \bmmu}$ denote the marginal distribution of $Y_d$ under $P_{\bmmu}$. Denote the variance of $Y_d$ under a distribution that generates the data (the data-generating process) by $\sigma^2_d$. Let $\calF_t = \sigma(D_1, Y_1, \ldots, D_t, Y_t)$ be the sigma-algebra. We denote the absolute expected gap between the outcomes of the binary treatments by $\Delta_\bmmu \coloneqq \abs{\mu_1 - \mu_0}$. 
In the bandit problem, this gap plays an important role in theoretical evaluations. 

\subsection{Contributions}
We design a minimax and Bayes optimal adaptive experiment for treatment choice. Our proposed experimental design utilizes the Neyman allocation and an empirical best treatment recommendation. The Neyman allocation is a classical allocation rule that allocates treatments in the ratio of the standard deviations \citep{Neyman1934ontwo}. Since we do not know the standard deviations in advance, directly implementing the Neyman allocation is infeasible. To address this problem, we split the treatment allocation phase into two stages. In the first stage, we allocate each treatment with the same ratio. After the first stage, using the observations, we estimate the standard deviations. In the second stage, we allocate treatments so that the empirical allocation ratio matches the estimated standard deviation. After the Neyman allocation, we estimate the expected outcomes using the sample mean and recommend the treatment with the highest sample mean as the best treatment. 

Surprisingly, this single design attains both minimax and Bayes optimality; that is, the lower and upper bounds exactly match, including the constant terms. Let $\calA$ be the set of possible experiments $\p{\p{D_t^{\delta}}_{t \in [T]}, \widehat{d}_T^{\delta}}$. We also let $\delta^{\text{TSNA}}$ be our proposed TSNA experiment. Here, for simplicity, we assume that the potential outcomes follow Gaussian distributions with variances $\sigma^2_1$ and $\sigma^2_0$, which will be generalized in the main sections. Then, we show the following results about the minimax and Bayes optimality:
\begin{align*}
    \text{(Minimax optimality)}\ \ \ \ \ &\limsup_{T \to \infty} \sup_{\bmmu \in \calM^2} \sqrt{T}\text{Regret}^{\delta^{\text{TSNA}}}_{\bmmu}\\
    &\leq \bigp{\sigma_1 + \sigma_0}\Phi(-1)\\
    &\leq \inf_{\delta \in \calA}\liminf_{T \to \infty} \sup_{\bmmu \in \calM^2} \sqrt{T}\text{Regret}^\delta_{\bmmu},\\
    \text{(Bayes optimality)}\ \ \ \ \ \ &\limsup_{T \to \infty} T\int_{\bmmu \in \calM^2} \text{Regret}^{\delta^{\text{TSNA}}}_{\bmmu}\rmd H(\bmmu)\\
    &\leq \frac{1}{4}\sum_{d\in\{1, 0\}}\int_{\mu \in \calM}h_d(\mu \mid \mu_{\backslash d})\bigp{\sigma_1 + \sigma_0}^2\rmd H_{\backslash d}(\mu)\\
    &\leq \inf_{\delta \in \calA}\liminf_{T \to \infty} T\int_{\bmmu \in \calM^2} \text{Regret}^\delta_{\bmmu}\rmd H(\bmmu),
\end{align*}
where ``$\backslash \{d\}$'' denotes the treatment $b \in \{1, 0\} \backslash \{d\}$, and $h_d(\mu\mid \mu_{\backslash d})$ is the positive continuous derivative of $H_d(\mu) \coloneqq \bbP_{H}\bigp{\mu_d \leq \mu\mid \mu_{\backslash d}}$.

\section{Literature review}
\label{sec:review}
Treatment choice and experimental design have long been studied, influencing each other. 
The origin of the treatment choice problem dates back to statistical decision making \citep{Wald1949statisticaldecision,Haavelmo1944theprobability}. In the 1940s, experimental design was also extensively studied \citep{Thompson1933onthe}. For example, \citet{Wald1945sequentialtests} develops sequential testing methods. After those works, the problems of treatment choice and experimental design have been investigated using various approaches. One of the popular approaches for these problems is multi-armed bandits, where we aim to design adaptive experiments to conduct decision making about treatments \citep{Gittins1989multiarmed}. 

\paragraph{Multi-armed bandits.}
Among various formulations in multi-armed bandits, cumulative reward maximization and BAI are two of the main interests. In cumulative reward maximization, we aim to maximize the sum of the outcomes obtained in an experiment. The regret in this setting is different from the one in this study and is called the in-sample regret in some academic fields. In contrast, BAI, an instance of pure exploration, attempts to identify the best treatment after the experiment, with the aim of implementing the treatment for the future population. The earliest BAI formulation appeared under the name \emph{ordinal optimization} \citep{Chen2000simulationbudget,Glynn2004largedeviations}, focusing on non-adaptive optimal designs via large-deviation principles.  
That literature often assumes that an experimenter knows how to allocate treatments to attain optimality, which requires knowledge of the distributional information of the treatments' outcomes. 
Beginning in the 2010s, BAI was formulated by explicitly addressing the estimation of the optimal allocation rule \citep{Audibert2010bestarm,Bubeck2011pureexploration}. BAI is typically studied under two settings: the fixed-confidence setting and the fixed-budget setting. In the fixed-confidence setting, we first fix a target error probability, while the sample size $T$ is left unspecified. Treatments are sampled until the probability of misidentification is theoretically guaranteed to be below a pre-specified threshold. On the other hand, fixed-budget BAI aims to minimize the misidentification probability $\bbP_{\bmmu}\p{\widehat{d}^\delta_T \neq d^*_{\bmmu}}$ or the regret $\text{Regret}^\delta_{\bmP_{\bmmu}}$ given a fixed sample size $T$. Our study focuses solely on the fixed-budget setting and we refer to it simply as BAI.

\paragraph{Multi-armed bandits and treatment choice.}
In the 2000s, the treatment choice problem also started to attract attention. \citet{Manski2002treatmentchoice} introduces statistical decision making to causal inference and establishes the treatment choice framework. Multi-armed bandits and treatment choice problems essentially treat the same problems, although the former mainly focuses on the experimental design, while the latter focuses on statistical analysis of decision making using observations.

\paragraph{Regret analysis.} Regret analysis is a core interest in theoretical analysis in multi-armed bandits and treatment choice. The regret is decomposed as follows:
\[
\text{Regret}^\delta_{\bmP_{\bmmu}} = \Delta_{\bmmu} \cdot \bbP_{\bmmu} \p{\widehat{d}^\delta_T \neq d^*_{\bmmu}},
\]
where recall that $\Delta_\bmmu = \abs{\mu_1 - \mu_0}$ denotes the gap in expected outcomes between the binary treatments, which is the same as the average treatment effect (ATE).  

The optimality in terms of the simple regret depends on how we deal with uncertainty about the underlying distribution $P_{\bmmu}$. There are mainly the following three types of evaluation frameworks:
\begin{itemize}
    \item \textbf{Distribution-dependent analysis:} Evaluate performance under a fixed distribution $P_{\bmmu}$. 
    \item \textbf{Minimax analysis:}  Evaluate performance under the worst case of $P_{\bmmu}$ among a set of distributions $\calP$.
    \item \textbf{Bayesian analysis:} Evaluate performance by averaging over $P_{\bmmu}$ weighted by a prior. 
\end{itemize}
Under distribution-dependent analysis for BAI, the misidentification probability $\bbP_{\bmmu}\p{\widehat{d}^\delta_T \neq d^*_{\bmmu}}$ decays at an \emph{exponential} rate in $T$. Since $\Delta_{\bmmu}$ is fixed, it can be asymptotically ignored. That is, we evaluate the regret by the convergence rate of $\bbP_{\bmmu}\p{\widehat{d}^\delta_T \neq d^*_{\bmmu}}$ because
\[\frac{1}{T}\log \text{Regret}^\delta_{\bmmu} \approx \frac{1}{T}\log \bbP_{\bmmu}\p{\widehat{d}^\delta_T \neq d^*_{\bmmu}}\]
holds for large $T$. Lower bounds for this probability have been developed by \citet{Kaufmann2014complexity,Kaufmann2016complexity}. For binary treatments with Gaussian outcomes and known variances, \citet{Kaufmann2014complexity,Kaufmann2016complexity} show that Neyman allocation is optimal, which allocates treatments in proportion to their standard deviations. They also show that when outcomes follow a one-parameter exponential family and the number of treatments is two, uniform allocation is nearly optimal. When variances are unknown, \citet{Kato2025neymanallocation} proves that for binary treatments, Neyman allocation with adaptive variance estimation remains optimal in a local regime where $\Delta_\bmmu$ is small. This is because the estimation error of the standard deviation affects the evaluation of the probability of misidentification. 

This study focuses on minimax and Bayesian frameworks. Under these frameworks, distributions with parameters such that $\Delta_\bmmu = C/\sqrt{T}$ for some constant $C > 0$ dominate the regret, unlike the pointwise analysis with $\mu_1 - \mu_0$ fixed for $T$. We briefly overview this mechanism. First, it is known that the probability of misidentification has the following exponential convergence to zero:
\[\bbP_{\bmmu} \p{\widehat{d}^\delta_T \neq d^*_{\bmmu}} = O\p{\exp \p{- C T \Delta^2_\bmmu}}.\]
for some constant $C > 0$. 
Using this bound, we obtain the following regret decomposition:
\[
\text{Regret}^\delta_{\bmmu} = \Delta_\bmmu \cdot \bbP_{\bmmu} \p{\widehat{d}^\delta_T \neq d^*_{\bmmu}} \leq  O\p{\Delta_\bmmu \cdot \exp \p{- C T \Delta^2_\bmmu}}.
\]
From this bound, we observe the following cases:
\begin{itemize}
    \item If $\Delta_\bmmu$ converges to zero at a rate slower than $1/\sqrt{T}$, then there exists a function $g(T) \to \infty$ such that $\text{Regret}^\delta_{\bmmu} = O\bigp{\exp(-g(T))}$.
    \item If $\Delta_\bmmu = C_1/\sqrt{T}$ for some constant $C_1 > 0$, then the regret behaves as $\text{Regret}^\delta_{\bmmu} = O\p{\frac{1}{\sqrt{T}}}$. This follows because $\exp(- C T \Delta^2_{\bmmu}) = \exp(- C C_1^2)$ becomes constant in $T$.
    \item If $\Delta_\bmmu$ converges to zero at a rate faster than $1/\sqrt{T}$, then $\text{Regret}^\delta_{\bmmu} = o(1/\sqrt{T})$ holds.
\end{itemize}
Therefore, instances $P_{\bmmu}$ with parameters $\bmmu$ such that $\Delta_\bmmu = C_1 / \sqrt{T}$ dominate the regret in the worst case. In Bayesian analyses, this worst case dominates the regret bound. Therefore, it is sufficient to consider such local alternatives when deriving minimax or Bayes lower bounds.

Minimax rate-optimal BAI experiments are developed in \citet{Bubeck2011pureexploration}, whereas Bayes rate-optimal BAI experiments are proposed by \citet{Komiyama2023rateoptimal}. These results achieve optimal convergence rates, but exact constant matching between upper and lower bounds remains unresolved in general. Our contribution addresses this gap. We derive tight minimax and Bayes lower bounds, including exact constants, and propose a single adaptive design whose simple regret asymptotically attains these bounds. 

\paragraph{Change-of-measure arguments.} 
Our lower bound depends on the arguments in theorems in \citet{Kaufmann2016complexity}, which develops tight lower bounds for BAI algorithms utilizing the change-of-measure arguments. The change-of-measure arguments are a strong tool for deriving lower bounds and have been applied in various statistical problems, including cumulative reward maximization \citep{Lai1985asymptoticallyefficient}, nonparametric regression \citep{Tsybakov2008introductionnonparametric}, and the limit-of-experiment framework \citep{LeCam1972theoryofstatisics,LeCam1986asymptoticmethods,VanderVaart1991anasymptotic}. 

In the treatment choice problem, the change-of-measure arguments have been used for asymptotic analysis in the form of the limit-of-experiment framework \citep{Hirano2009asymptotics}. Although some studies claim that the lower bounds in bandits and the limit-of-experiment framework are different \citep{Adusumilli2023risk}, we claim that they are based on essentially the same tool. Therefore, we can develop lower bounds following similar arguments in other bandit studies. In this study, we develop lower bounds using the change-of-measure arguments and discuss that they are based on the same tool as in the limit-of-experiment framework but use fewer restrictions. For example, we do not use Le Cam's third lemma and the asymptotic representation theorems, which are unnecessary for discussing the asymptotic optimality in the treatment choice problem, although we use tools similar to them with fewer assumptions. 

In this sense, we contribute to a broad literature of the efficiency bounds arguments for adaptive experimental design in causal inference, discussed in recent studies in econometrics such as \citet{Armstrong2022asymptoticefficiency} and \citet{Hirano2025asymptoticrepresentations}. We further claim that we do not have to use diffusion processes to derive optimal algorithms in adaptive experimental design for treatment choice, unlike the results in \citet{Adusumilli2023risk}.

\paragraph{Adaptive experimental design for efficient ATE estimation.}
Adaptive experimental design for efficient ATE estimation has long been studied \citep{vanderLaan2008theconstruction,Hahn2011adaptiveexperimental,Kato2020efficientadaptive,Kato2024activeadaptive}. From the algorithmic level, our experiment resembles the two-stage experiment for efficient average treatment effect estimation proposed in \citet{Hahn2011adaptiveexperimental}. In fact, treatment choice and ATE estimation are closely related. For example, if we consider experiments using some estimators of the expected outcome $\widehat{\mu}_d$ and recommend the treatment with the highest estimated expected outcome ($\widehat{d}_T = \argmax_{d\in\{1, 0\}}\widehat{\mu}_d$), from Markov's inequality, we can upper bound the regret as follows:
\[\text{Regret}^\delta_{\bmP_{\bmmu}} =  \Delta_\bmmu\cdot \bbP_{\bmmu} \p{\widehat{d}^\delta_T \neq d^*_{\bmmu}} \leq \bbE\sqb{\abs{\bigp{\widehat{\mu}_1 - \widehat{\mu}_0} - \bigp{\mu_1 - \mu_0}}},\]
where $\mu_1 - \mu_0$ is the ATE. Thus, adaptive experimental design for efficient estimation of the ATE intuitively yields optimal experiments for treatment choice. 

In ATE estimation, several works propose sequential estimation of the ideal allocation ratio \citep{Kato2020efficientadaptive,Cook2024semiparametricefficient,Dai2023clipogdogd,Neopane2024logarithmicneyman,Noarov2025strongerneyman}. Sequential estimation improves finite-sample performance in ATE estimation, and we expect that these results can be applied in our setting, which is an important direction for future work.

\section{Two-stage Neyman allocation}
\label{sec:tsna}
This section defines our proposed adaptive experiment, the TSNA experiment. The TSNA experiment aims to allocate treatment in the ratio of the standard deviations and recommend the treatment with the highest sample mean at the end of the experiment. Since the experimenter does not know the standard deviations in advance, the experimenter estimates them during the experiment. We realize this approach as follows: the experimenter splits the treatment allocation phase into two stages, where the first stage consists of $\ceil{rT}$ rounds and the second stage of $T - \ceil{rT}$ rounds for some constant $r \in (0, 1)$ independent of $T$. Note that as $T \to \infty$, both $rT$ and $(1 - r)T$ diverge to infinity. For simplicity, the experimenter chooses $r$ so that $\ceil{rT / 2} \in [2, T - 1]$. In the first stage, the experimenter allocates treatments uniformly across all treatments, allocating each treatment $rT / 2$ rounds for experimental subjects. After the first stage, the experimenter estimates the standard deviations using the observations. In the second stage, the experimenter allocates each treatment so that the empirical treatment allocation ratio matches the estimated standard deviations. After the second stage of the treatment allocation phase, the experimenter goes into the treatment choice phase. In this phase, the experimenter recommends the treatment with the highest sample mean as the best treatment. Allocation based on the standard deviations is referred to as the Neyman allocation. Therefore, 
we refer to this experiment as the TSNA experiment and denote it by $\delta^{\text{TSNA}}$. In Algorithm~\ref{alg:tsNA}, we show the pseudo-code. 

\begin{algorithm}[t!]
\caption{TSNA experiment $\delta^{\text{TSNA}}$}
\label{alg:tsNA}
\begin{algorithmic}[1]
\STATE Total horizon $T$, and split ratio $r\in(0,1)$. 
\STATE \textbf{Treatment allocation phase.}
\STATE \textbf{First stage: uniform allocation}. 
    \FOR{$t = 1$ \textbf{to} $\ceil{rT / 2}$}
        \STATE Allocate treatment $D_t = 1$.
        \STATE Observe the outcome $Y_t$.
    \ENDFOR
    \FOR{$t = \ceil{rT / 2} + 1$ \textbf{to} $\ceil{rT}$}
        \STATE Allocate treatment $D_t = 0$.
        \STATE Observe the outcome $Y_t$.
    \ENDFOR
\STATE \textbf{Second stage: allocation using an estimated standard deviations}. 
\STATE Estimate an ideal treatment allocation ratio $w^*$, as defined in \eqref{eq:estoptprob}. 
\FOR{$t = \ceil{rT}+1$ \textbf{to} $T$}
    \STATE Allocate $D_t$ following the Bernoulli distribution with parameter $\widehat{\pi}_{rT}$. 
    \STATE Observe $Y_t$
\ENDFOR
%--------------------------------------------------
\STATE \textbf{Treatment choice phase}.
\STATE $\widehat{d}^{\delta^{\text{TSNA}}}_T = \argmax_d\widehat{\mu}_{d,T}$. 
\end{algorithmic}
\end{algorithm}

\subsection{Treatment allocation phase: the two-stage rule}
\label{sec:allocationphase}
We aim to allocate treatments so that the empirical allocation ratio $\frac{1}{T}\sum_{t=1}^T \mathbbm{1}[D_t = 1]$ converges to the allocation ratio $w^*$ defined as
\begin{align*}
    w^* &\coloneqq \frac{\sigma_1}{\sigma_1 + \sigma_0},
\end{align*}
where recall that $\sigma^2_d$ denotes the variance of the outcome under the data-generating process. We refer to $w^*$ as the ideal allocation ratio, which corresponds to the Neyman allocation. In many studies such as \citet{Garivier2016optimalbest}, this ideal allocation ratio is used as a target in the treatment allocation phase of adaptive experimental design. Since the variances are unknown, the ideal allocation ratio is also unknown. Therefore, using estimates of the standard deviations, we estimate the ideal allocation ratio $w^*$.

To estimate the standard deviations, we split the treatment allocation phase into two stages. In the first stage, each treatment is allocated an equal number of times, that is, $rT / 2$ rounds per treatment. Based on the outcomes, we estimate the standard deviations of the potential outcomes $Y_1$ and $Y_0$. In the second stage, we allocate treatments proportionally to the estimated standard deviations. We describe the experimental procedure in detail below.

\textbf{First stage.}
We allocate each treatment $\ceil{rT / 2}$ rounds. Specifically, for $t = 1,2,\dots, \ceil{rT / 2}$, we allocate treatment $1$, and for $t = \ceil{rT / 2} + 1, \ceil{rT / 2} + 2, \dots, \ceil{rT}$, we allocate treatment $0$. 

\textbf{Second stage.}
The allocation in the second stage depends on the estimated standard deviations. 

For simplicity, let $\ceil{rT / 2}$ be an integer. For each treatment $d\in \{1, 0\}$, we estimate the variance $\sigma^2_d$ of $Y_d$ as
\begin{align*}
    \widehat{\sigma}^2_{d, rT} &\coloneqq \frac{1}{rT / 2 - 1} \sum_{s=1}^{rT} \mathbbm{1}[D_s = d] \Bigp{Y_s - \widehat{\mu}_{d, rT}}^2,
\end{align*}
where $\widehat{\mu}_{d, rT} \coloneqq \frac{1}{rT / 2}\sum_{s=1}^{rT} \mathbbm{1}[D_s = d] Y_s$.
Here, we use the unbiased sample variance, but one can also use the sample variance that divides the sum by $rT / 2$ instead of $rT / 2 - 1$. Then, we estimate the ideal allocation ratio as
\begin{align}
\label{eq:estoptprob}
    \widehat{w}_{rT} &\coloneqq \frac{ \widehat{\sigma}_{1, rT}}{ \widehat{\sigma}_{1, rT}  +  \widehat{\sigma}_{0, rT}}.
\end{align}
We then allocate treatments in the second stage by sampling $D_t \in \{1, 0\}$ from a Bernoulli distribution with parameter $\widehat{\pi}_{rT}$, which is defined as
\begin{align}
\label{eq:estallocation}
    \widehat{\pi}_{rT} \coloneqq \frac{\widetilde{\pi}_{1, rT}}{\widetilde{\pi}_{1, rT} + \widetilde{\pi}_{0, rT}},
\end{align}
where $\widetilde{\pi}_{1, rT} \coloneqq \max\cb{\widehat{w}_{rT} - \frac{r}{(1 - r)2}, 0}$ and $\widetilde{\pi}_{0, rT} \coloneqq \max\cb{1 - \widehat{w}_{rT} - \frac{r}{(1 - r)2}, 0}$. That is, we allocate treatment $1$ with probability $\widehat{\pi}_{rT}$ and treatment $0$ with probability $1 - \widehat{\pi}_{rT}$.

\subsection{Treatment choice phase}
After observing $Y_T$, we recommend the treatment with the highest sample mean as the estimate of the best treatment; that is, 
\[
\widehat{d}^{\delta^{\text{TSNA}}}_T \coloneqq \argmax_{d \in \{1, 0\}} \widehat{\mu}_{d, T}.
\]

\section{Outcome models}
\label{sec:parametricdist}
From this section onward, we present the theoretical analysis of our TSNA experiment, aiming to show its minimax and Bayes optimality. We first define a class of distributions $\calP$ for outcomes $Y$, over which we discuss the worst-case and average performance of experiments. 

As a distribution class, we use canonical exponential families, defined as
\[
\calP \coloneqq \cb{ (P_\theta)_{\theta \in \Theta}
\colon  
\frac{dP_\theta}{d\xi}(y) = \exp\bigp{ y \theta - b\bigp{\theta} }},
\]
where $P_\theta$ is a distribution parameterized by a natural parameter $\theta$ (not $P_{\mu}$ used in the other parts), $\Theta \subset \bbR$ is the space of natural parameters $\theta$, $\xi$ is some reference measure on $\calY$, and $b \colon \Theta \to \bbR$ is a convex and twice differentiable function. In this study, however, we consider worst-case and average performance in terms of the mean parameter and characterize the lower and upper bounds in terms of variances, where the mean corresponds to $\dot{b}(\theta)$ and the variance corresponds to $\ddot{b}(\theta)$. Therefore, it is more convenient to define a class of distributions based on the mean and variance parameters. 

This section provides such a definition and introduces an outcome model as a set of $2$ classes of distributions. Based on this motivation, we define the following mean-parameterized sub-Gaussian exponential families, which include Gaussian and Bernoulli distributions as special cases. This class is essentially the same as the standard canonical exponential family, but defined so that the inverse Fisher information becomes the variance.

\begin{definition}[Mean-parameterized canonical exponential family]
\label{def:mean_param}
Let $\xi$ be some reference measure on $\calY$. 
Let $\calM\subset\bbR$ be a non-empty compact interval,
and let $\sigma^2:\calM\to(0,\infty)$ be a twice continuously differentiable function.

Define $\calP(\sigma^2,\calM,\calY)$ to be the collection of all families $\{P_\mu:\mu\in\calM\}$ for which there exist:
\begin{itemize}
\item an open interval $\Theta\subset\bbR$ (natural-parameter space),
\item a strictly convex, three-times continuously differentiable log-partition function $b:\Theta\to\bbR$,
\item a continuously differentiable map $\theta:\calM\to\Theta$,
\end{itemize}
such that for every $\mu\in\calM$, the following holds:
\begin{enumerate}[leftmargin=*,itemsep=2pt,topsep=2pt,label=(\roman*)]
\item \textbf{Compactness:} $\overline{\theta(\calM)}\subset \Theta$. 
\item \textbf{Density:} $P_\mu\ll\xi$ with $\frac{dP_\mu}{d\xi}(y)=\exp\Bigp{y \theta(\mu)-b\big(\theta(\mu)\big)}$, 
and, for all $\theta\in\theta(\calM)$, $\int_{\calY}\exp\bigp{y\theta-b(\theta)} \rmd \xi(y)=1$ holds. 
\item \textbf{Mean-parameterization:} $\dot{b}\bigp{\theta(\mu)}=\mu$ for all $\mu\in\calM$ (equivalently, on $\theta(\calM)$ we have $\theta=\bigp{\dot{b}}^{-1}$).
\item \textbf{Prescribed variance:} $\ddot{b}\bigp{\theta(\mu)}=\sigma^2(\mu)$ for all $\mu\in\calM$.
\item \textbf{Sub-Gaussian distribution:} for all $\mu \in \calM$, $Y - \mu$ ($Y \sim P_{\mu}$) is sub-Gaussian.
\end{enumerate}
We call any such family a mean-parameterized canonical exponential family with variance $\sigma^2$.
\end{definition}

Notably, the following properties hold for the mean-parameterized canonical exponential family. 
\begin{proposition}
For any $P_\mu\in\calP(\sigma^2,\calM,\calY)$, the following holds:
\begin{enumerate}[topsep=0pt, itemsep=0pt, partopsep=0pt, leftmargin=*]
\renewcommand\labelenumi{(\theenumi)}
    \item \label{enu:2} For each $\mu \in \calM$, the Fisher information $I_\mu > 0$ of $P_\mu$ exists and is equal to the inverse of the variance $1/\sigma^2(\mu)$.
    \item \label{enu:3} Let $\ell_\mu(y) = \log f_\mu(y)$ be the likelihood function of $P_\mu$, and $\dot{\ell}_\mu$, $\ddot{\ell}_\mu$, and $\dddot{\ell}_\mu$ be the first, second, and third derivatives of $\ell_\mu$ for $\mu$. The likelihood function $\ell_\mu$ is three times differentiable and satisfies the following properties:
    \begin{enumerate}
        \item $\bbE_{\bmmu}\left[\dot{\ell}_\mu(Y)\right] = \bbE_{\bmmu}\left[\bigp{Y - \mu}/\sigma^2(\mu)\right] = 0$;
        \item $\bbE_{\bmmu}\left[\ddot{\ell}_\mu(Y)\right] = -I_\mu = - 1/\sigma^2(\mu)$;
        \item For each $\mu \in \calM$, there exist a neighborhood $U_\mu$ and a function $u_\mu(y) \geq 0$, and the following holds:
        \begin{enumerate}
            \item $\left|\ddot{\ell}_{\tau}(y)\right| \leq u_\mu(y)\ \ \ \forall\ \tau \in U_\mu$;
            \item $\bbE_{\bmmu}\left[u_\mu(Y)\right] < \infty$. 
        \end{enumerate}
    \end{enumerate}
\end{enumerate}
\end{proposition}

\begin{remark}
The outcome space $\calY$ and the parameter space $\calM$ should be carefully chosen to satisfy the conditions in Definition~\ref{def:mean_param}. For example, if the outcome $Y_d$ follows a Bernoulli distribution with support $\calY = \{0, 1\}$, we can choose $\calM$ as $\calM = [c, 1 - c]$, where $c > 0$ is some positive constant. If we instead choose $\calM = [0, 1]$, the variance becomes $0$ at $\mu = 0$ and $\mu = 1$, which violates our definition of the class. In this case, the Fisher information does not exist at $\mu = 0$ and $\mu = 1$, since the Fisher information is given by $I(\mu) = \frac{1}{\mu(1 - \mu)}$.
\end{remark}

For each $d\in\{1, 0\}$, let $\sigma^2_d:\calM \to (0, \infty)$ be a variance function that is continuous in $\mu \in \calM$. Then, given $\bm{\sigma}^2 \coloneqq (\sigma^2_d)_{d\in\{1, 0\}}, \calM, \calY$, we define an outcome model $\calB$ as the following set of distributions:
\[
\calB_{\bm{\sigma}^2} \coloneqq \Bigcb{(P_d)_{d\in\{1, 0\}}\colon \forall d\in \{1, 0\}\;\;P_d \in \calP(\sigma^2_d(\cdot), \calM, \calY)}.
\]
In other words, an element $\bmP$ in $\calB_{\bm{\sigma}^2}$ is a family of parametric distributions as defined in Definition~\ref{def:mean_param}; that is, $\bmP = (P_{d, \mu_d})_{d\in\{1, 0\}}$. When we emphasize the parameters, we denote the distribution by $\bmP_{\bmmu}  = (P_{d, \mu_d})_{d\in\{1, 0\}}$, where $\bmmu = (\mu_d)_{d\in\{1, 0\}}$.

\begin{example}[Distributions]
Our distribution class $\calB_{\bm{\sigma}^2}$ allows heterogeneity across treatment. Typical choices include:
\begin{enumerate}[topsep=2pt,itemsep=2pt,leftmargin=*,label=(\alph*)]
\item Mixed families for example, a Bernoulli distribution for treatment $1$ (variance $\mu_d(1 - \mu_d)$), and a Gaussian distribution for treatment $0$.
\item Homogeneous family with treatment–specific variance functions: for example, all treatments Negative Binomial with different $r_d$ giving $\sigma_d^2(\mu)=\mu+\mu^2/r_d$.
\end{enumerate}
\end{example}

\section{Lower bounds}
\label{sec:lowerbound}
In this section, we derive minimax and Bayes lower bounds. We first define a class of experiments for which these lower bounds hold and then present each of the minimax and Bayes lower bounds. 

\subsection{Minimax lower bound}
We now present the minimax lower bound, which characterizes the best possible performance in the worst-case distribution.

\begin{theorem}[Minimax lower bound]
\label{thm:minimax_lowerbound}
Let $\calA$ be the set of possible experiments $\p{\p{D_t^{\delta}}_{t \in [T]}, \widehat{d}_T^{\delta}}$. Fix an outcome space $\calY$, a parameter space $\calM \subset \bbR$, and a set of variance functions $\bm{\sigma}^2 = (\sigma^2_d)_{d \in \{1, 0\}}$ with $\sigma^2 \colon \{1, 0\} \times \calM \to (0, \infty)$. Suppose that the marginal distribution of each $Y_{d, t}$ is $P_{d, \mu_d}$ such that $\bmP_{\bmmu} = (P_{d, \mu_d})_{d\in\{1, 0\}} \in \calB_{\bm{\sigma}^2}$. Then the following lower bound holds:
\begin{align*}
    &\inf_{\delta \in \calA} \liminf_{T \to \infty} \sqrt{T} \sup_{\bmmu \in \calM^2} \text{Regret}^\delta_{\bmmu} \geq \Bigp{\overline{\sigma}_1 + \overline{\sigma}_0} \Phi(-1),
\end{align*}
where $\overline{\sigma}^2_d \coloneqq \sup_{\mu\in\calM} \sigma^2_d(\mu)$.
\end{theorem}
Here, the regret is scaled by $\sqrt{T}$, which reflects the convergence rate.

\subsection{Bayes lower bound}
We now derive a Bayes lower bound. Let $H$ be a prior distribution on $\calM^2$. We assume the following regularity conditions for the prior distribution:

\begin{assumption}[Uniform continuity of conditional densities]
\label{asm:uniformcontinuity}
There exist uniformly continuous conditional probability density functions $h_1(\mu_1 \mid \mu_2)$, $h_0(\mu_0 \mid \mu_1)$, and $h_{10}(\mu_1, \mu_0)$. That is, for every $\epsilon > 0$, there exists $\delta(\epsilon) > 0$ such that for all $\mu_d, \lambda_d \in \calM$ such that $|\mu_d - \lambda_d| \leq \delta(\epsilon)$, we have
    \[
    \left| h_1(\mu_1 \mid \mu_0) - h_1(\lambda_1 \mid \mu_0) \right| \leq \epsilon,\quad  \left| h_0(\mu_0 \mid \mu_1) - h_0(\lambda_0 \mid \mu_1) \right| \leq \epsilon.
    \]
\end{assumption}
This assumption follows those in Theorem~1 of \citet{Lai1987adaptivetreatment} and Assumption~1 of \citet{Komiyama2023rateoptimal}. 

For a prior $\Pi$ satisfying Assumption~\ref{asm:uniformcontinuity}, the following Bayes lower bound holds. 
\begin{theorem}[Bayes lower bound]
\label{thm:bayes_lowerbound}
Let $\calA$ be the set of possible experiments $\p{\p{D_t^{\delta}}_{t \in [T]}, \widehat{d}_T^{\delta}}$. Fix an outcome space $\calY$, a parameter space $\calM \subset \bbR$, and a set of variance functions $\bm{\sigma}^2 = (\sigma^2_d)_{d \in \{1, 0\}}$ with $\sigma^2_d \colon \calM \to (0, \infty)$. Suppose that the marginal distribution of each $Y_{d, t}$ is $P_{d, \mu_d}$ such that $\bmP_{\bmmu} = (P_{d, \mu_d})_{d\in\{1, 0\}} \in \calB_{\bm{\sigma}^2}$. Then, for any prior $H$ satisfying Assumption~\ref{asm:uniformcontinuity}, the following lower bound holds:
\[
\inf_{\delta \in \calA} \liminf_{T \to \infty} T \int_{\bmmu \in \calM^2} \text{Regret}^{\delta}_{\bmmu}  \rmd H(\bmmu)
\geq \frac{1}{4}\sum_{d\in\{1, 0\}}\int_{\mu \in \calM}h_d(\mu \mid \mu_{\backslash d})\bigp{\sigma_1(\mu) + \sigma_0(\mu)}^2\rmd H_{\backslash d}(\mu),
\]
where $\backslash \{d\}$ denotes the treatment $b \in \{1, 0\} \backslash d$.
\end{theorem}

\subsection{Approach to the proof}
This section explains our approach to the proof. For the detailed proof, see Appendices~\ref{appdx:proofminimaxlowerbound} and \ref{appdx:proofBayeslowerbound}.
We develop our lower bounds by utilizing the change-of-measure arguments, which are a standard approach in statistical analysis. For example, the change-of-measure arguments have been applied to the derivation of lower bounds in cumulative reward maximization in bandits \citep{Lai1985asymptoticallyefficient}, nonparametric regression \citep{Stone1980optimalrates}, and the limit-of-experiment framework \citep{VanderVaart1998asymptoticstatistics}. Among various results in the change-of-measure literature, we use the following lemma proved in \citet{Kaufmann2016complexity}. 

\begin{lemma}[Lemma~18 in \citet{Kaufmann2016complexity}]
\label{lem:changeofmeasure}
For $\bmmu, \bmnu \in \calM^2$, let $\bmP_{\bmmu}$ and $\bmP_{\bmnu}$ be two bandit
models such that for all $d \in \{1, 0\}$, the marginal distributions $P_{d, \mu_d}$ and $P_{d, \nu_d}$ of $Y_d$ are mutually absolutely continuous. 
Then, for every $E_T \in \calF_T$, it holds that
\[
\bbP_{\bmmu}\bigp{E_T} = \bbE_{\bmnu}\sqb{\mathbbm{1}[E_T] \exp\bigp{ - L_T}},
\]
where
\[L_T \coloneqq \sum_{d\in\{1, 0\}}\sum^T_{t=1}\mathbbm{1}[D_t = d]\log\p{\frac{f_{\nu_d}(Y_t)}{f_{\mu_d}(Y_t)}},\]
where recall that $f(\cdot; \mu)$ is the probability density function of $P_{d, \mu_d}$.
\end{lemma}

This lemma is quite general. Although it is originally developed for deriving distribution-dependent lower bounds in BAI, it can also be applied to deriving lower bounds in cumulative reward maximization. We can also interpret this result as sharing a similar motivation to Le Cam's third lemma, but it is more appropriately designed for treatment choice. In this study, we use this lemma to derive minimax and Bayes lower bounds.

To apply the change-of-measure lemma to the treatment choice problem, we use the following lemma. 
\begin{lemma}
\label{lem:lan}
    For every $h_1, h_0 \in \bbR$ such that $\mu_d = \nu_d + h_d / \sqrt{T} \in \calM$, it holds that
    \begin{align*}
        &\sum_{d\in\{1, 0\}}\sum^T_{t=1}\mathbbm{1}[D_t = d]\log\p{\frac{f_{\nu_d}(Y_t)}{f_{\nu_d + h_d / \sqrt{T}}(Y_t)}}\\
        &= - \frac{1}{\sqrt{T}}\sum^T_{t=1}\mathbbm{1}[D_t = d]h_d \dot{\ell}_{\nu_d}(Y_{d, t}) + \frac{h^2_d}{2\sigma^2(\nu_d)}\frac{1}{T}\sum^T_{t=1}\bbE_{\bmnu}\bigsqb{\mathbbm{1}[D_t = d]} + o_{\bmP_{\bmnu}}\p{1}.
    \end{align*}
\end{lemma}
This lemma is an extension of Theorem~7.2 in \citet{VanderVaart1998asymptoticstatistics}. Since the observations are dependent, we apply the martingale law of large numbers instead of the law of large numbers for i.i.d.\ samples used in the original theorem of \citet{VanderVaart1998asymptoticstatistics}.

In the minimax analysis, we develop lower bounds as follows. First, we decompose the worst-case regret as
\begin{align*}
    \sup_{\bmmu \in \calM^2} \text{Regret}^\delta_{\bmmu}
    &= \max_{d^\dagger\in\{1, 0\}}\sup_{\bmmu\colon \argmax_{d\in\{1, 0\}} \mu_d = d^\dagger} \Delta_{\bmmu} \cdot \bbP_{\bmmu} \p{\widehat{d}^\delta_T \neq d^*_{\bmmu}}.
\end{align*}
We first consider the case where $d^\dagger = 1$. We restrict the baseline distribution $\bmP_{\bmnu}$ to a distribution whose parameter $\bmnu = (\nu_d)_{d\in\{1, 0\}}$ is given as
\begin{align*}
\nu_1 = \nu_0 = \mu.
\end{align*}
We restrict the alternative distribution $\bmP_{\bmnu}$ to a distribution whose parameter $\bmmu = (\mu_d)_{d\in \{1, 0\}}$ is given as
\begin{align*}
   \mu_1 = \mu + h_1 / \sqrt{T},\quad \mu_0 = \mu - h_0 / \sqrt{T},
\end{align*}
where
\begin{align*}
    h_1 &\coloneqq \sigma^2_1(\mu) \Big/\sqrt{V\p{\overline{D}_T}},\qquad h_2 \coloneqq \sigma^2_0(\mu) \Big/\sqrt{V\p{\overline{D}_T}},\\
    V\p{\overline{D}_T} &\coloneqq \frac{\sigma^2_1(\mu)}{\overline{D}_T} + \frac{\sigma^2_0(\mu)}{1 - \overline{D}_T}\\
    \overline{D}_T &\coloneqq \frac{1}{T}\sum^T_{t=1}\bbE_{\bmnu}\bigsqb{\mathbbm{1}[D_t = d]}.
\end{align*}
From Lemmas~\ref{lem:changeofmeasure} and \ref{lem:lan}, and the uniform integrability of $\exp\p{-\lambda\dot{\ell}_{\mu}(Y_{d, t})}$, for some $\lambda > 0$ and an event $E_{1, T}$ such that $\widehat{d}_T \neq d^\dagger = 1$, it holds that
\begin{align*}
    &\bbP_{\bmmu}(E_{1, T}) =\\
    &\bbE_{\bmnu}\sqb{\mathbbm{1}[E_{1, T}]\exp\p{\sum_{d\in\{1,0\}}\frac{1}{\sqrt{T}}\sum^T_{t=1}\mathbbm{1}[D_t = d]h_d \dot{\ell}_{\nu_d}(Y_{d, t}) - \sum_{d\in\{1,0\}}\frac{h^2_d}{2\sigma^2(\nu_d)}\frac{1}{T}\sum^T_{t=1}\bbE_{\bmnu}\bigsqb{\mathbbm{1}[D_t = d]} + o(h^2_d) }}.
\end{align*}
Similarly, in the case where $d^\dagger = 0$, for an event $E_{0, T}$ such that $\widehat{d}_T \neq d^\dagger = 1$, by setting the alternative parameters as 
\begin{align*}
   \mu_1 = \mu - h_1,\quad \mu_0 = \mu + h_0,
\end{align*}
we have
\begin{align*}
    &\bbP_{\bmmu}(E_{0, T}) =\\
    &\bbE_{\bmnu}\sqb{\mathbbm{1}[E_{0, T}]\exp\p{- \sum_{d\in\{1,0\}}\frac{1}{\sqrt{T}}\sum^T_{t=1}\mathbbm{1}[D_t = d]h_d \dot{\ell}_{\nu_d}(Y_{d, t}) - \sum_{d\in\{1,0\}}\frac{h^2_d}{2\sigma^2_d(\nu_d)}\frac{1}{T}\sum^T_{t=1}\bbE_{\bmnu}\bigsqb{\mathbbm{1}[D_t = d]} + o(h^2_d) }}.
\end{align*}
Here, we have
\begin{align*}
    \sup_{\bmmu \in \calM^2} \text{Regret}^\delta_{\bmmu}
    &= \bigabs{h_1 - h_0}\inf_{E_{1,T},E_{0,T}}\max\bigcb{\bbP_{\bmmu}(E_{1, T}), \bbP_{\bmmu}(E_{0, T})}.
\end{align*}
By computing this expression, we can obtain the minimax lower bound. We can also develop Bayes regret in a similar way. 

\section{Upper bounds and asymptotic optimality}
\label{sec:upperbound}
In this section, we establish an upper bound for the simple regret for the TSNA experiment. The performance upper bound depends on the parameters of the distributions. By taking the worst case for the parameters, we can develop the worst-case upper bound. In addition, by taking the average of the upper bound weighted by the prior distribution, we can develop the average upper bound. 

We also demonstrate that these worst-case and average upper bounds match the minimax and Bayes lower bounds derived in Section~\ref{sec:lowerbound}. Therefore, we can conclude that our proposed experiment is asymptotically minimax and Bayes optimal. 

\subsection{The worst-case upper bound and minimax optimality}
First, we derive the following worst-case upper bound for the simple regret under the TSNA experiment. The proof is shown in Appendix~\ref{appdx:proof:worstcase_upper}.
\begin{theorem}
\label{thm:minimax_upperbound}
Fix an outcome space $\calY$, a parameter space $\calM \subset \bbR$, and a set of variance functions $\bm{\sigma}^2 = \bigp{\sigma^2_d}_{d\in\{1, 0\}}$, where $\sigma^2_d \calM \to (0, \infty)$. Let $\overline{\sigma}^2_d \coloneqq \sup_{\mu\in\calM} \sigma^2_d(\mu)$. 
Suppose that the marginal distribution of each $Y_{d, t}$ is $P_{d, \mu_d}$ such that $\bmP_{\bmmu} = (P_{d, \mu_d})_{d\in\{1, 0\}} \in \calB_{\bm{\sigma}^2}$. If
$r / 2 \leq \min_{d\in\{1, 0\}}\overline{\sigma}_d \big/ \sum_{b \in \{1, 0\}} \overline{\sigma}_b$, then the regret of the TSNA experiment satisfies
    \begin{align*}
    &\limsup_{T \to \infty} \sup_{\bmmu \in \calM^2} \sqrt{T} \text{Regret}^{\delta^{\text{TSNA}}}_{\bmmu} \leq 
     \Bigp{\overline{\sigma}_1 + \overline{\sigma}_0}\Phi(-1).
\end{align*}
\end{theorem}
Thus, we upper bound the simple regret of the proposed experiment in Theorem~\ref{thm:minimax_upperbound}. 

The results in the minimax lower bound (Theorem~\ref{thm:minimax_lowerbound}) and the worst-case upper bound (Theorem~\ref{thm:minimax_upperbound}) imply the asymptotic minimax optimality.

\begin{corollary}[Asymptotic minimax optimality]
    Under the same conditions in Theorems~\ref{thm:minimax_lowerbound} and \ref{thm:minimax_upperbound}, it holds that
    \begin{align*}
        &\limsup_{T \to \infty} \sup_{\bmmu \in \calM^2} \sqrt{T} \text{Regret}^{\delta^{\text{TSNA}}}_{\bmmu}\\
        &\leq 
     \Bigp{\overline{\sigma}_1 + \overline{\sigma}_0}\Phi(-1)\\
        &\leq \inf_{\delta \in \calA} \liminf_{T \to \infty} \sqrt{T} \sup_{\bmmu \in \calM^2} \text{Regret}^\delta_{\bmmu}.
    \end{align*}
    Thus, the TSNA experiment is asymptotically minimax optimal. 
\end{corollary}

Note that our asymptotic minimax optimality does not restrict the distribution to be local ones, which has been considered in existing studies \citep{Kato2024generalizedneyman,Kato2025neymanallocation,Hirano2025asymptoticrepresentations,Armstrong2022asymptoticefficiency,Adusumilli2022neymanallocation,Adusumilli2023risk}. We point out that localization appears as a global optimum because the global worst case is characterized by $1/\sqrt{T}$.

\subsection{The average upper bound and Bayes optimality}
Next, we derive the following average upper bound for the expected simple regret under the TSNA experiment. The proof is shown in Appendix~\ref{appdx:proof:average_upper}.

\begin{theorem}[Average upper bound]
\label{thm:bayes_upperbound}
Fix an outcome space $\calY$, a parameter space $\calM \subset \bbR$, and a set of variance functions $\bm{\sigma}^2 = \bigp{\sigma^2_d}_{d\in\{1, 0\}}$, where $\sigma^2_d \times \calM \to (0, \infty)$. Suppose that the marginal distribution of each $Y_{d, t}$ is $P_{d, \mu_d}$ such that $\bmP_{\bmmu} = (P_{d, \mu_d})_{d\in\{1, 0\}} \in \calB_{\bm{\sigma}^2}$. Also suppose that $r / 2 \leq \min_{a\neq b}\sigma_d(\mu) \big/ \bigp{\sigma_d(\mu) + \sigma_b(\mu)}$ holds for all $\mu\in\calM$. Then, the TSNA experiment satisfies the following average upper bound:
    \begin{align*}
        &\limsup_{T \to \infty} T\int_{\bmmu \in \calM^2} \text{Regret}^{\delta^{\text{TSNA}}}_{\bmmu}\rmd H(\bmmu)\\
    &\ \ \ \ \ \ \ \leq \frac{1}{4}\sum_{d\in\{1, 0\}}\int_{\mu \in \calM}h_d(\mu \mid \mu_{\backslash d})\bigp{\sigma_1(\mu) + \sigma_0(\mu)}^2\rmd H_{\backslash d}(\mu).
\end{align*}
\end{theorem}

The results in the Bayes lower bound (Theorem~\ref{thm:bayes_lowerbound}) and the average upper bound (Theorem~\ref{thm:bayes_upperbound}) imply the asymptotic Bayes optimality.

\begin{corollary}[Asymptotic Bayes optimality]
    Under the same conditions in Theorems~\ref{thm:bayes_lowerbound} and \ref{thm:bayes_upperbound}, as $r\to 0$, it holds that
    \begin{align*}
        &\limsup_{T \to \infty} T\int_{\bmmu \in \calM^2} \text{Regret}^{\delta^{\text{TSNA}}}_{\bmmu}\rmd H(\bmmu)\\
        &\ \ \ \ \ \ \ \leq \frac{1}{4}\sum_{d\in\{1, 0\}}\int_{\mu \in \calM}h_d(\mu \mid \mu_{\backslash d})\bigp{\sigma_1(\mu) + \sigma_0(\mu)}^2\rmd H_{\backslash d}(\mu)\\
        &\ \ \ \ \ \ \ \leq \inf_{\delta \in \calA}\liminf_{T \to \infty} T\int_{\bmmu \in \calM^2} \text{Regret}^{\delta}_{\bmmu}\rmd H(\bmmu).
    \end{align*}
    Thus, the TSNA experiment is asymptotically Bayes optimal. 
\end{corollary}

\subsection{Approach to the proof}
As well as the proof for the lower bounds, our proofs are based on the following regret decomposition:
\[
\text{Regret}^\delta_{\bmmu} = \Delta_{\bmmu} \bbP_{\bmmu} \p{\widehat{d}^{\delta^{\text{TSNA}}}_t \neq d^*_{\bmmu}},
\]
which implies that the upper bound of
\[
\bbP_{\bmmu} \p{\widehat{d}^{\delta^{\text{TSNA}}}_t \neq d^*_{\bmmu}}
\]
plays an important role in the derivation of the upper bound on the simple regret.

In the proof, we utilize both large deviation bounds and the central limit theorem. For an adaptive experiment similar to ours, 
Theorem~1 in \citet{Hahn2011adaptiveexperimental} establishes the asymptotic normality of the ATE estimator as follows.
\begin{proposition}[Theorem~1 in \citet{Hahn2011adaptiveexperimental}]
\label{prp:hahnclt}
et $w \in (0, 1)$ be a value such that 
\[\frac{1}{T}\sum^T_{t=1}\mathbbm{1}[D_t = 1] \xrightarrow{\text{a.s.}} w.\]
Suppose that, for all $d\in \{1, 0\}$, $w \in (0, 1)$ holds and $w$ is independent of $T$. Then, under $P_{\bmmu}$ and the TSNA experiment, for each $d \neq d^*_{\bmmu}$, we have
\begin{align*}
    \sqrt{T}\p{\Bigp{\widehat{\mu}_{d^*_{\bmmu}, T} - \widehat{\mu}_{d, T}} - \Bigp{\mu_{d^*_{\bmmu}} - \mu_d}} \xrightarrow{\rmd} \calN(0, V)\quad (T \to \infty),
\end{align*}
where 
\[V\p{w^*} \coloneqq \frac{\sigma^2_1(\mu_1)}{w^*} + \frac{\sigma^2_0(\mu_0)}{1 - w^*}.\]
\end{proposition}
The central limit approximation gives us the following probability evaluation:
\[\lim_{T\to\infty}\bbP_{\bmmu}\p{\sqrt{T}\p{\Bigp{\widehat{\mu}_{d^*_{\bmmu}, T} - \widehat{\mu}_{d, T}} - \Bigp{\mu_{d^*_{\bmmu}} - \mu_d}} \leq C} = \Phi\p{\frac{C}{\sqrt{V\p{w^*}}}},\]
where $C$ is independent of $T$. Therefore, when $\sqrt{T}(\mu_1 - \mu_0) = O(1)$ ($T \to \infty$), we can approximate the misidentification probability. 

When $\sqrt{T}(\mu_1 - \mu_0) \to \infty$ ($T \to \infty$), the central limit approximation cannot be applied. In that region, we use large deviation bounds to evaluate the misidentification probability. We show the large deviation bound in Lemma~\ref{lem:upper_probabilityofmisidentiifcation} as 
\[\bbP_{\bmmu}\p{\widehat{\mu}_{d^*_{\bmmu}, T} \leq \widehat{\mu}_{d, t}} \leq 2\exp\p{-\frac{rT \Delta^2_{\bmmu}}{16 v}},\]
for $v > 0$ is some constant independent of $T$, corresponding to the variance proxy of the potential outcome distribution. 
To proof this upper bound, we apply the Chernoff bound. 

Since the regret is decomposed as $\text{Regret}^\delta_{\bmmu} = \Delta_{\bmmu} \cdot \bbP_{\bmmu} \p{\widehat{d}^\delta_T \neq d^*_{\bmmu}}$, by using the central limit bound and the large deviation bound, we can obtain an upper bound on the regret. Then, by taking the worst case, we derive the worst-case regret, and by taking the average over the prior, we derive the average regret. 

\section{On the ideal allocation ratio}
We define the ideal allocation ratio as
\begin{align*}
    w^* &\coloneqq \frac{\sigma_1}{\sigma_1 + \sigma_0},
\end{align*}
where $\sigma_d$ denotes the standard deviation of the data-generating process. However, as our theory suggests, in the worst-case analysis it is enough to consider the allocation ratio under the worst-case standard deviation, that is,
\begin{align*}
    w^{\text{worst}\mathchar`-\text{case}} &\coloneqq \frac{\overline{\sigma}_1}{\overline{\sigma}_1 + \overline{\sigma}_0};
\end{align*}
For the average analysis, it is enough to consider the allocation ratio when $\sigma^2_1(\mu) = \sigma^2_0(\mu)$ for some $\mu\in\calM$, that is,
\begin{align*}
    w^{\text{average}} &\coloneqq \frac{\overline{\sigma}_1(\mu)}{\overline{\sigma}_1(\mu) + \overline{\sigma}_0(\mu)} = 1/2
\end{align*}
for some $\mu\in\calM$. This fact suggests simplifications in specific cases. For example, when outcomes follow a Bernoulli distribution, our experiment can be simplified by omitting the variance estimation step. When dealing with a Bernoulli distribution, the worst-case and average allocations reduce to the uniform allocation, that is, $w^* \approx 1/2$.

\section{Conclusion}
We designed a minimax and Bayes optimal adaptive experiment for treatment choice. Our designed experiment is based on the Neyman allocation; that is, an experimenter allocates treatments in the ratio of the standard deviations. Since the standard deviations are unknown, we estimate them at the first stage of the treatment allocation phase and use these estimates at the second stage. We show that this single design surprisingly attains minimax and Bayes optimality. Furthermore, we developed a theoretical framework that is simpler than the limit-of-experiment framework but yields stronger implications. Moreover, the optimality of our design does not depend on the specific behavior of the experiment, such as the convergence of the treatment allocation. Our result contributes not only to adaptive experimental design but also to the entire theoretical analysis of treatment choice problems.

\bibliography{arXiv2.bbl}

\bibliographystyle{tmlr}

\clearpage 

\onecolumn

\appendix

\tableofcontents

\clearpage 

\section{Proof of minimax lower bounds (Theorem~\ref{thm:minimax_lowerbound})}
\label{appdx:proofminimaxlowerbound}
This section presents the proof of the minimax lower bounds.

\subsection{Proof of the minimax lower bound (Proof of Theorem~\ref{thm:minimax_lowerbound})}

\begin{proof}
We decompose the worst-case regret as
\begin{align*}
    \sup_{\bmmu \in \calM^2} \text{Regret}^\delta_{\bmmu} 
    &= \max_{d^\dagger\in\{1, 0\}}\sup_{\bmmu\colon \argmax_{d\in\{1, 0\}} \mu_d = d^\dagger} \text{Regret}^\delta_{\bmmu}\\
    &= \max_{d^\dagger\in\{1, 0\}}\sup_{\bmmu\colon \argmax_{d\in\{1, 0\}} \mu_d = d^\dagger} \Delta_{\bmmu} \cdot \bbP_{\bmmu} \p{\widehat{d}^\delta_T \neq d^*_{\bmmu}}.
\end{align*}

We first investigate the case where $d^\dagger = 1$. To derive a lower bound, we utilize the change-of-measure arguments, which develop lower bounds using two different distributions. We refer to $\bmP_{\bmm}$ as the alternative distribution and $\bmP_{\bmnu}$ as the baseline distribution. 

From Lemma~18 in \citet{Kaufmann2016complexity} (see Lemma~\ref{lem:changeofmeasure} in our Section~\ref{sec:lowerbound}), for any event $E \in \calF_T$, and for mutually absolutely continuous $\bmP_{\bmmu}$ and $\bmP_{\bmnu}$, it holds that
\[\bbP_{\bmmu}(E) = \bbE_{\bmnu}\Bigsqb{\mathbbm{1}[E]\exp\bigp{- L_T}},\]
where 
\[L_T \coloneqq \sum_{d\in\{1, 0\}}\sum^T_{t=1}\mathbbm{1}[D_t = d]\log\p{\frac{f_{\nu_d}(Y_t)}{f_{\mu_d}(Y_t)}}.\]
We derive lower bounds based on this result.

First, we specify the parameter $\bmmu \in \calM^2$ of the alternative distribution $\bmP_{\bmmu}$ and the parameter $\bmnu \in \calM^2$ of the baseline distribution $\bmP_{\bmnu}$. Fix $\mu \in \calM$. We restrict the baseline distribution $\bmP_{\bmnu}$ to a distribution whose parameter $\bmnu = (\nu_d)_{d\in\{1, 0\}}$ is given as
\begin{align*}
\nu_1 = \nu_0 = \mu.
\end{align*}
We restrict the alternative distribution $\bmP_{\bmnu}$ to a distribution whose parameter $\bmmu = (\mu_d)_{d\in \{1, 0\}}$ is given as
\begin{align*}
   \mu_1 = \mu + h_1,\quad \mu_0 = \mu - h_0,
\end{align*}
where
\begin{align*}
    h_1 &\coloneqq \sigma^2_1(\mu) \Big/\sqrt{V\p{\overline{D}_T}},\qquad h_2 \coloneqq \sigma^2_0(\mu) \Big/\sqrt{V\p{\overline{D}_T}},\\
    V &\coloneqq \frac{\sigma^2_1(\mu)}{\overline{D}_T} + \frac{\sigma^2_0(\mu)}{1 - \overline{D}_T}\\
    \overline{D}_T &\coloneqq \frac{1}{T}\sum^T_{t=1}\bbE_{\bmnu}\bigsqb{\mathbbm{1}[D_t = d]}.
\end{align*}

From Lemma~\ref{lem:lan}, we have
\begin{align*}
    L_T &= \sum_{d\in\{1, 0\}}\sum^T_{t=1}\mathbbm{1}[D_t = d]\log\p{\frac{f_{\nu_d}(Y_t)}{f_{\nu_d + h_d}(Y_t)}}\\
    &= - G_T + Z_T + o_{\bmP_{\bmnu}}(1),
\end{align*}
where 
\begin{align*}
    G_T &\coloneqq \frac{1}{\sqrt{T}}\sum^T_{t=1}\p{\mathbbm{1}[D_t = 1]h_1 \dot{\ell}_{\mu}(Y_{1, t}) - \mathbbm{1}[D_t = 0]h_0 \dot{\ell}_{\mu}(Y_{0, t})} + o_{\bmP_{\bmnu}}\bigp{1},\\
    Z_T &\coloneqq \frac{h^2_1}{2\sigma^2_1(\mu)}\overline{D}_T + \frac{h^2_0}{2\sigma^2_0(\mu)}\Bigp{1 - \overline{D}_T}.
\end{align*}
Here, it holds that $Z_T = \frac{1}{2}$. Also note that $G_T$ converges in distribution to the standard normal distribution from the martingale central limit theorem. 

Let $\calE_{1, T} = \cb{\widehat{d}_T \neq 1} \subset \calF_T$. Then, for every $E_{1, T} \in \calE_{1, T}$, we have
\[\bbP_{\bmmu}\bigp{E_{1, T}} = \bbE_{\bmnu}\Bigsqb{\mathbbm{1}[E_{1, T}]\exp\bigp{- L_T}} = \bbE_{\bmnu}\Bigsqb{\mathbbm{1}[E_{1, T}]\exp\p{G_T - \frac{1}{2}}}.\]

Similarly, in the case where $d^\dagger = 0$, we set the parameter of the alternative distribution as
\begin{align*}
   \mu_1 = \mu - h_1,\quad \mu_0 = \mu + h_0.
\end{align*}
We use the same baseline distribution as in the case where $d^\dagger = 1$; that is, $\nu_1 = \nu_0 = \mu$. Let $\calE_{0, T} = \cb{\widehat{d}_T \neq 0} \subset \calF_T$. Then, we have
\[\bbP_{\bmmu}\bigp{E_{0, T}} = \bbE_{\bmnu}\sqb{\mathbbm{1}[E_{0, T}]\exp\p{- G_T - \frac{1}{2}}}.\]

By combining them, we have
\begin{align*}
    &\sup_{\bmmu \in \calM^2}\text{Regret}^\delta_{\bmmu}\\
    &\geq \max_{d^\dagger \in \{1, 0\}}\sup_{\bmmu \in \calM^2\colon \argmax_{d\in\{1, 0\}}\mu_d = d^\dagger}\text{Regret}^\delta_{\bmmu}\\
    &\geq \inf_{E_{1,T} \in \calE_{1, T},E_{0,T} \in \calE_{0, T}}\max_{d^\dagger \in \{1, 0\}}\sup_{\bmmu \in \calM^2\colon \argmax_{d\in\{1, 0\}}\mu_d = d^\dagger}\Delta_{\bmmu} \cdot \bbP_{\bmmu}\bigp{E_d}\\
    &\geq \bigabs{h_1 - h_0}\\
    &\ \ \ \cdot \inf_{E_{1,T} \in \calE_{1, T},E_{0,T} \in \calE_{0, T}}\max\cb{
    \bbE_{\bmnu}\sqb{\mathbbm{1}\sqb{E_{1,T}} \exp\p{G_T - \frac{1}{2}}},\bbE_{\bmnu}\sqb{\mathbbm{1}\sqb{E_{0,T}} \exp\p{- G_T - \frac{1}{2}}}}.
\end{align*}

Let us define
\[g_T \coloneqq \inf_{E_{1,T} \in \calE_{1, T},E_{0,T} \in \calE_{0, T}}\max\cb{
    \bbE_{\bmnu}\sqb{\mathbbm{1}\sqb{E_{1,T}} \exp\p{G_T - \frac{1}{2}}},\bbE_{\bmnu}\sqb{\mathbbm{1}\sqb{E_{0,T}} \exp\p{- G_T - \frac{1}{2}}}},\]
and consider the lower bound of $g_T$. Fix $T$ and suppress the subscript $T$ for notational convenience.
Let $G$ be the random variable on $(\Omega,\calF,\bbP_{\bmnu})$.
The events $E_1,E_0$ are a partition of $\Omega$, under which
$E_1\cap E_0=\varnothing$ and
$E_1\cup E_0=\Omega$ hold. Set
\[
I_1 \coloneqq \mathbbm{1}\sqb{E_1},\qquad I_0 \coloneqq \mathbbm{1}\sqb{E_0} = 1 - I_1.
\]
Let us also define
\begin{align*}
    A &\coloneqq \bbE_{\bmnu}\sqb{I_1 \cdot \exp\p{G - \frac{1}{2}}},\\
    B &\coloneqq \bbE_{\bmnu}\sqb{I_0 \cdot \exp\p{- G - \frac{1}{2}}}.   
\end{align*}

For every $\omega \in \Omega$ we have $I_1(\omega), I_0(\omega) \in \{0,1\}$
and $I_1(\omega) + I_0(\omega) = 1$, hence
\begin{align*}
    &I_1(\omega) \exp\bigp{G(\omega)} + I_0(\omega) \exp\bigp{-G(\omega)}\\
    &\in \Bigcb{\exp\bigp{G(\omega)}, \exp\bigp{-G(\omega)}}\\
    &\geq \min\Bigcb{\exp\bigp{G(\omega)}, \exp\bigp{-G(\omega)}}\\
    &= \exp\Bigp{-|G(\omega)|}.
\end{align*}
Multiplying by $\exp\bigp{-1/2}$ and taking expectations gives
\begin{align*}
A + B
&= \bbE\sqb{I_1 \exp\p{G-\frac{1}{2}} + I_0 \exp\p{-G-\frac{1}{2}}} \\
&= \exp\p{-\frac{1}{2}}\bbE\Bigsqb{I_1 \exp\bigp{G} + I_0 \exp\bigp{-G}} \\
&\geq \exp\p{-\frac{1}{2}}\bbE\Bigsqb{\exp\bigp{-|G|}}.
\end{align*}
Therefore, it holds that
\[
\max\{A,B\}\geq \frac{A+B}{2}
\geq \exp\p{-\frac{1}{2}}\frac{1}{2}\bbE\Bigsqb{\exp\bigp{-|G|}}.
\]

Since this holds for every partition $(E_1,E_0)$, we obtain
for each $T$:
\begin{align}
g_T \geq
\exp\p{-\frac{1}{2}}\frac{1}{2}\bbE_{\bmnu} \Bigsqb{\exp\bigp{-|G_T|}}.
\label{eq:vn-lower}
\end{align}

The function $q(x) \coloneqq \exp\bigp{-|x|}$ is uniformly integrable and continuous on $\bbR$. Let $N$ be a random variable that follows the standard normal distribution.
Since $G_T$ converges in distribution to the standard normal distribution, we have
\[
\bbE_{\bmnu}\Bigsqb{\exp\bigp{-|G_T|}}\rightarrow \bbE_{\bmnu}\Bigsqb{\exp\bigp{-|N|}}\quad (T\to\infty).
\]

We now compute $\bbE[\exp\bigp{-|N|}]$ explicitly. Using symmetry of $N$,
\[
\bbE\bigl[\exp\bigp{-|N|}\bigr]
=
2 \int_0^\infty \exp\bigp{-x} \phi(x)\rmd x,
\]
where $\phi(x) = (2\pi)^{-1/2} \exp\bigp{-x^2/2}$ is the standard normal density.
We have
\begin{align*}
\int_0^\infty \exp\bigp{-x} \phi(x)\rmd x
&= \frac{1}{\sqrt{2\pi}}\int_0^\infty \exp\bigp{-x - x^2/2}\rmd x \\
&= \frac{1}{\sqrt{2\pi}}\int_0^\infty \exp\bigp{-(x^2+2x)/2}\rmd x \\
&= \frac{1}{\sqrt{2\pi}}\int_0^\infty \exp\bigp{-(x+1)^2/2 + 1/2}\rmd x \\
&= \frac{\exp\bigp{1/2}}{\sqrt{2\pi}}\int_0^\infty \exp\bigp{-(x+1)^2/2}\rmd x.
\end{align*}
With the substitution $w = x+1$ we obtain
\begin{align*}
\int_0^\infty \exp\bigp{-x} \phi(x)\rmd x
&= \frac{\exp\bigp{1/2}}{\sqrt{2\pi}}\int_1^\infty \exp\bigp{-w^2/2}\rmd y \\
&= \exp\p{\frac{1}{2}}\int_1^\infty \phi(w)\rmd y\\
&= \exp\p{\frac{1}{2}}\bigp{1 - \Phi(1)}\\
&= \exp\p{\frac{1}{2}}\Phi(-1).
\end{align*}
Thus, we have
\[
\bbE\bigsqb{\exp\bigp{-|N|}} = 2 \exp\p{\frac{1}{2}}\Phi(-1),
\qquad
\frac12\bbE\bigsqb{\exp\bigp{-|N|}} = \exp\p{\frac{1}{2}}\Phi(-1).
\]

Combining this with \eqref{eq:vn-lower} yields
\[
\liminf_{n\to\infty} g_T
\geq
\exp\p{-\frac{1}{2}}\frac{1}{2}\bbE\bigl[\exp\bigp{-|N|}\bigr]
=
\exp\p{-\frac{1}{2}} \exp\p{\frac{1}{2}}\Phi(-1)
=
\Phi(-1).
\]
Hence
\begin{align}
\liminf_{n\to\infty} g_T \geq \Phi(-1).
\label{eq:liminf}
\end{align}

Recall that 
\begin{align}
    h_1 - h_0 = \sqrt{\frac{V\p{\overline{D}_T}}{T}}.
\end{align}
Therefore, we have
\begin{align*}
    &\sqrt{T}\sup_{\bmmu \in \calM^2}\text{Regret}^\delta_{\bmmu} \geq \sqrt{V\p{\overline{D}_T}}g_T.
\end{align*}
Hence, it holds that
\begin{align*}
    &\liminf_{T\to\infty}\sqrt{T}\sup_{\bmmu \in \calM^2}\text{Regret}^\delta_{\bmmu}\geq \sqrt{V\p{\overline{D}_T}}\Phi(-1) = \sqrt{\frac{\sigma^2_1(\mu)}{\overline{D}_T} + \frac{\sigma^2_0(\mu)}{1 - \overline{D}_T}}\Phi(-1).
\end{align*}
Lastly, by taking the infimum for $\overline{D}_T$, we obtain
\begin{align*}
    &\inf_{w\in(0, 1)}\liminf_{T\to\infty}\sqrt{T}\sup_{\bmmu \in \calM^2}\text{Regret}^\delta_{\bmmu}\geq \inf_{w\in(0, 1)} \sqrt{\frac{\sigma^2_1(\mu)}{w} + \frac{\sigma^2_0(\mu)}{1 - w}}\Phi(-1).
\end{align*}
The optimal solution for $w$ is given as 
\[w^* \coloneqq \frac{\sigma_1(\mu)}{\sigma_1(\mu) + \sigma_0(\mu)}.\]
Then, for each $\mu\in\calM$, we have
\begin{align*}
    &\inf_{w\in(0, 1)}\liminf_{T\to\infty}\sqrt{T}\sup_{\bmmu \in \calM^2}\text{Regret}^\delta_{\bmmu}\geq \sqrt{V\p{\overline{D}_T}}\Phi(-1) = \Bigp{\sigma_1(\mu) + \sigma_0(\mu)}\Phi(-1).
\end{align*}
By taking the supremum for $\mu \in \calM$, we have
\begin{align*}
    &\inf_{w\in(0, 1)}\liminf_{T\to\infty}\sqrt{T}\sup_{\bmmu \in \calM^2}\text{Regret}^\delta_{\bmmu}\geq \sup_{\mu \in \calM}\sqrt{V\p{\overline{D}_T}}\Phi(-1) = \Bigp{\sigma_1(\mu) + \sigma_0(\mu)}\Phi(-1).
\end{align*}
This completes the proof. 
\end{proof}

\section{Proof of Bayes lower bounds (Theorem~\ref{thm:bayes_lowerbound})}
\label{appdx:proofBayeslowerbound}
\begin{proof}[Proof of Theorem~\ref{thm:bayes_lowerbound}]
The Bayes regret is decomposed as
\begin{align*}
    \int_{\bmmu \in \calM^2} \text{Regret}^\delta_{\bmmu} \rmd H(\bmmu) = \int_{\bmmu \in \calM^2} \Bigp{\mu_{d^*_{\bmmu}} - \bbE_{\bmmu}\sqb{\mu_{\widehat{d}^{\delta^{\text{TSNA}}}_T}}} \rmd H(\bmmu).
\end{align*}
For $\bmmu \in \calM^2$, let $d^{*(m)}_{\bmmu}$ be the index of the $m$-th largest element. For example, $d^{*(1)}_{\bmmu} = d^*_{\bmmu}$. Let us also define the following set of parameters:
\begin{align*}
    \Lambda_d &\coloneqq \cb{\bmmu \in \calM^2 \colon d^*_{\bmmu} = d}.
\end{align*}
Then, the following holds:
\begin{align*}
    &\int_{\bmmu \in \calM^2} \Bigp{\mu_{d^*_{\bmmu}} - \bbE_{\bmmu}\sqb{\mu_{\widehat{d}^{\delta^{\text{TSNA}}}_T}}} \rmd H(\bmmu)\\
    &= \sum_{d^\dagger\in\{1, 0\}} \int_{\bmmu \in \calM^2} \mathbbm{1}\bigsqb{\bmmu \in \Lambda_d} \Bigp{\mu_{d^\dagger} - \mu_{d^{(2)}_\bmmu}} \bbP_{\bmmu}\p{\widehat{d}^{\delta^{\text{TSNA}}}_T \neq d^\dagger}\rmd H(\bmmu).
\end{align*}

Let $\bmnu = (\nu_d)_{d\in\{1, 0\}}$ be the parameter of the baseline distribution and define them as
\begin{align*}
    \nu_1 = \nu_0 = \widetilde{m},
\end{align*}
where 
\[\widetilde{m} = \frac{\sigma_0(\mu_1)\mu_1 + \sigma_1(\mu_1)\mu_0}{\sigma_1(\mu_1) + \sigma_0(\mu_0)}.\]

Consider the case where $d^\dagger = 1$. Then, we consider a lower bound of 
\[\int_{\bmmu \in \calM^2} \mathbbm{1}\bigsqb{\bmmu \in \Lambda_d} \Bigp{\mu_1 - \mu_0} \bbP_{\bmmu}\p{\widehat{d}^{\delta^{\text{TSNA}}}_T \neq 1}\rmd H(\bmmu).\]
Let us consider restricting the set $\calM^2$ of parameters $\bmmu$ as
\[\widetilde{\calM}(h, T) \coloneqq \cb{\bmmu \in \calM^2 \colon \mu_1 - \mu_2 = \frac{h}{\sqrt{T}}},\]
where $h \in \bbR$ is independent of $T$. Let $\calH$ be the space of $h_1 - h_0$. 

From Lemma~\ref{lem:lan}, we have
\begin{align*}
    L_T &= \sum_{d\in\{1, 0\}}\sum^T_{t=1}\mathbbm{1}[D_t = d]\log\p{\frac{f_{\nu_d}(Y_t)}{f_{\nu_d + h_d}(Y_t)}}\\
    &= - G_T + Z_T + o_{\bmP_{\bmnu}}(1),
\end{align*}
where 
\begin{align*}
    G_T &\coloneqq \frac{\mu_1 - \mu_0}{\sigma_1(\mu_1) + \sigma_0(\mu_0)}\frac{1}{\sqrt{T}}\sum^T_{t=1}\p{\sigma_1(\mu_1)\mathbbm{1}[D_t = 1] \dot{\ell}_{\mu}(Y_{1, t}) - \sigma_0(\mu_0)\mathbbm{1}[D_t = 0]h_0 \dot{\ell}_{\mu}(Y_{0, t})} + o_{\bmP_{\bmnu}}(1),\\
    Z_T &\coloneqq \frac{\bigp{\mu_1 - \mu_0}^2}{2\bigp{\sigma_1(\mu_1) + \sigma_0(\mu_0)}^2}.
\end{align*}
Here, let $G_T = \frac{\mu_1 - \mu_0}{\sigma_1(\mu_1) + \sigma_0(\mu_0)} H_T$, and  
\[H_T \coloneqq \frac{1}{\sqrt{T}}\sum^T_{t=1}\p{\sigma_1(\mu_1)\mathbbm{1}[D_t = 1] \dot{\ell}_{\mu}(Y_{1, t}) - \sigma_0(\mu_0)\mathbbm{1}[D_t = 0]h_0 \dot{\ell}_{\mu}(Y_{0, t})}\]
is the term that converges to the standard normal distribution from the martingale central limit theorem. 

Let $\calE_{1, T} = \cb{\widehat{d}_T \neq 1} \subset \calF_T$. Then, for every $E_{1, T} \in \calE_{1, T}$, we have
\[\bbP_{\bmmu}\bigp{E_{1, T}} = \bbE_{\bmnu}\Bigsqb{\mathbbm{1}[E_{1, T}]\exp\bigp{- L_T}} \to \bbE_{\bmnu}\Bigsqb{\mathbbm{1}[E_{1, T}]\exp\bigp{G_T - Z_T}},\]
as $T\to\infty$. 

Similarly, we consider the case where $d^\dagger = 0$. We use the same baseline distribution as in the case where $d^\dagger = 1$; that is, $\nu_1 = \nu_0 = \overline{m}$. Let $\calE_{0, T} = \cb{\widehat{d}_T \neq 0} \subset \calF_T$. Then, we have
\[\bbP_{\bmmu}\bigp{E_{0, T}} \to \bbE_{\bmnu}\sqb{\mathbbm{1}[E_{0, T}]\exp\p{- G_T - Z_T}},\]
as $T\to\infty$. 

Let us define
\[U(\bmmu) \coloneqq \frac{\sqrt{T}\bigp{\mu_1 - \mu_0}}{\sigma_1(\mu_1) + \sigma_0(\mu_0)}.\]
Then, for large $T$, using the same argument as in the proof of Theorem~\ref{thm:minimax_lowerbound}, we have
\begin{align*}
    &\int_{\bmmu \in \calM^2} \Bigp{\mu_{d^*_{\bmmu}} - \bbE_{\bmmu}\sqb{\mu_{\widehat{d}^{\delta^{\text{TSNA}}}_T}}} \rmd H(\bmmu)\\
    &= \sum_{d^\dagger\in\{1, 0\}} \int_{\bmmu \in \calM^2} \mathbbm{1}\bigsqb{\bmmu \in \Lambda_d} \Bigp{\mu_{d^\dagger} - \mu_{d^{(2)}_\bmmu}} \bbP_{\bmmu}\p{\widehat{d}^{\delta^{\text{TSNA}}}_T \neq d^\dagger}\rmd H(\bmmu)\\
    &\geq \inf_{E_{1,T} \in \calE_{1, T},E_{0,T} \in \calE_{0, T}}\max \Biggcb{\int_{\bmmu \in \widetilde{\calM}^2(h, T)}\mathbbm{1}\bigsqb{\bmmu \in \Lambda_1}\bigp{\mu_1 - \mu_0}
    \bbE_{\bmnu}\sqb{\mathbbm{1}\sqb{E_{1,T}} \exp\p{U H_T - \frac{U^2}{2}}}\rmd H(\bmmu),\\
    &\ \ \ \ \ \ \ \ \ \ \ \ \ \ \ \ \ \ \ \ \ \ \ \ \ \ \ \ \ \ \ \ \ \ \ \ \ \ \ \int_{\bmmu \in \widetilde{\calM}^2(h, T)}\mathbbm{1}\bigsqb{\bmmu \in \Lambda_0}(\mu_0 - \mu_1)\bbE_{\bmnu}\sqb{\mathbbm{1}\sqb{E_{0,T}} \exp\p{- UH_T - \frac{U^2}{2}}}\rmd H(\bmmu)}\\
    &\geq \frac{1}{2}\Biggp{\int_{\bmmu \in \widetilde{\calM}^2(h, T)}\mathbbm{1}\bigsqb{\bmmu \in \Lambda_1}\Bigp{\mu_1 - \mu_0} \bbE_{\bmnu}\sqb{\mathbbm{1}\sqb{E_{1,T}} \exp\p{U H_T - \frac{U^2}{2}}}\rmd H(\bmmu)\\
    &\ \ \ \ \ \ \ \ \ \ \ \ \ \ \ \ \ \ \ \ \ \ \ \ \ \ \ \ \ \ \ \ \ \ \ \ \ \ \  + \int_{\bmmu \in \widetilde{\calM}^2(h, T)}\mathbbm{1}\bigsqb{\bmmu \in \Lambda_0}\Bigp{\mu_0 - \mu_1} \bbE_{\bmnu}\sqb{\mathbbm{1}\sqb{E_{0,T}} \exp\p{- UH_T - \frac{U^2}{2}}}\rmd H(\bmmu)},\\
    &\geq \int_{\bmmu \in \widetilde{\calM}^2(h, T)} \big|\mu_1 - \mu_0\big|\Phi\bigp{-U(\bmmu)} \rmd H(\bmmu).
\end{align*}

We have
\begin{align*}
    &\int_{\bmmu \in \widetilde{\calM}^2(h, T)} \big|\mu_1 - \mu_0\big|\Phi\bigp{-U(\bmmu)} \rmd H(\bmmu)\\
    &= \int_{\bmmu \in \widetilde{\calM}^2(h, T)}\mathbbm{1}\bigsqb{\mu_1 \geq \mu_0} \big|\mu_1 - \mu_0\big|\Phi\bigp{-U(\bmmu)} \rmd H(\bmmu) + \int_{\bmmu \in \widetilde{\calM}^2(h, T)}\mathbbm{1}\bigsqb{\mu_1 < \mu_0} \big|\mu_1 - \mu_0\big|\Phi\bigp{-U(\bmmu)} \rmd H(\bmmu)\\
    &= \int_{\mu_0 \in \calM}\int^{\mu_0 + h/\sqrt{T}}_{\mu_0} \big|\mu_1 - \mu_0\big|\Phi\bigp{-U(\bmmu)} \rmd H(\bmmu) + \int_{\mu_1 \in \calM}\int^{\mu_1 + h/\sqrt{T}}_{\mu_1}\big|\mu_1 - \mu_0\big|\Phi\bigp{-U(\bmmu)} \rmd H(\bmmu).
\end{align*}

From the uniform continuity of the prior (Assumption~\ref{asm:uniformcontinuity}), it holds that
\begin{align*}
    &\int_{\mu_0 \in \calM}\int^{\mu_0 + h/\sqrt{T}}_{\mu_0} \big|\mu_1 - \mu_0\big|\Phi\bigp{-U(\bmmu)} \rmd H(\bmmu)\\
    &= \int_{\mu_0 \in \calM}\int^{\mu_0 + h/\sqrt{T}}_{\mu_0} \big|\mu_1 - \mu_0\big|\Phi\bigp{-U(\bmmu)} h_1(\mu_1 \mid \mu_0) \rmd \mu_1 \rmd H_0(\mu_0)\\
    &\geq (1 - o(1))\int_{\mu_0 \in \calM}\int^{\mu_0 + h/\sqrt{T}}_{\mu_0} \big|\mu_1 - \mu_0\big|\Phi\bigp{-U(\bmmu)} h_1(\mu_0 \mid \mu_0) \rmd \mu_1 \rmd H_0(\mu_0)\\
    &= (1 - o(1))\int_{\mu_0 \in \calM}h_1(\mu_0 \mid \mu_0) \int^{\mu_0 + h/\sqrt{T}}_{\mu_0} \bigp{\mu_1 - \mu_0}\Phi\bigp{-U(\bmmu)}\rmd \mu_1 \rmd H_0(\mu_0)\\
    &\geq (1 - o(1))\int_{\mu_0 \in \calM}h_1(\mu_0 \mid \mu_0) \int^{\mu_0 + h/\sqrt{T}}_{\mu_0} \bigp{\mu_1 - \mu_0}\Phi\bigp{-\overline{U}(\bmmu)}\rmd \mu_1 \rmd H_0(\mu_0),
\end{align*}

On the local set $\{\mu_1 : \mu_1 \in [\mu_0, \mu_0 + h/\sqrt{T}]\}$ the difference $\mu_1 - \mu_0$ is of order $T^{-1/2}$, so we can approximate $\sigma_1(\mu_1)$ by $\sigma_1(\mu_0)$ in the denominator of $U$. Therefore, we have
\begin{align*}
    &(1 - o(1))\int_{\mu_0 \in \calM}h_1(\mu_0 \mid \mu_0) \int^{\mu_0 + h/\sqrt{T}}_{\mu_0} \bigp{\mu_1 - \mu_0}\Phi\bigp{-U(\bmmu)}\rmd \mu_1 \rmd H_0(\mu_0)\\
    &= (1 - o(1))\int_{\mu_0 \in \calM}h_1(\mu_0 \mid \mu_0) \int^{\mu_0 + h/\sqrt{T}}_{\mu_0} \bigp{\mu_1 - \mu_0}\Phi\bigp{-\overline{U}(\bmmu)}\rmd \mu_1 \rmd H_0(\mu_0),
\end{align*}
where
\[\overline{U}(\bmmu) \coloneqq \frac{\sqrt{T}\bigp{\mu_1 - \mu_0}}{\sigma_1(\mu_0) + \sigma_0(\mu_0)}.\]

We compute
\[
I_T \coloneqq \int_{\mu_0}^{\mu_0 + \frac{h}{\sqrt{T}}}
(\mu_1 - \mu_0)
\Phi\p{-\frac{\sqrt{T}(\mu_1 - \mu_0)}{\sigma_1(\mu_0) + \sigma_0(\mu_0)}}\rmd\mu_1.
\]
Let us define
\[C \coloneqq \sigma_1(\mu_0) + \sigma_0(\mu_0).\]
We now perform the change of variables as
\[
x = \frac{\sqrt{T}}{C}(\mu_1 - \mu_0)
\quad\Longleftrightarrow\quad
\mu_1 - \mu_0 = \frac{C}{\sqrt{T}}x,\qquad
\rmd\mu_1 = \frac{C}{\sqrt{T}}\rmd x.
\]
When $\mu_1 = \mu_0$, we have $x = 0$, and when
$\mu_1 = \mu_0 + h/\sqrt{T}$, we have $x = h/C$. Hence
\[
I_T
=
\int_{0}^{h/C}
\frac{C}{\sqrt{T}}x
\Phi(-x)
\frac{C}{\sqrt{T}}\rmd x
=
\frac{C^2}{T}\int_{0}^{h/C} x\Phi(-x)\rmd x.
\]

It remains to compute the definite integral
\[
J(a) \coloneqq \int_{0}^{a} x\Phi(-x)\rmd x,
\qquad a > 0.
\]
Recall that $\Phi(-x) = 1 - \Phi(x)$ and let $\phi(x)$ denote the standard
normal density. Using integration by parts with
\[
u = \Phi(-x),\quad \rmd v = x\rmd x,
\]
we have
\[
\rmd u = -\phi(x)\rmd x,\qquad v = \frac{x^2}{2},
\]
and therefore
\[
J(a)
= \left.\frac{x^2}{2}\Phi(-x)\right|_{x=0}^{x=a}
  + \frac{1}{2}\int_0^a x^2\phi(x)\rmd x.
\]
Since $\Phi(0)=1/2$, this gives
\[
J(a)
= \frac{a^2}{2}\Phi(-a) + \frac{1}{2}\int_0^a x^2\phi(x)\rmd x.
\]

To compute $\int_0^a x^2\phi(x)\rmd x$, note that
\[
\frac{\rmd}{\rmd x}\phi(x) = -x\phi(x),
\qquad
\frac{\rmd}{\rmd x}\bigl(x\phi(x)\bigr) = \phi(x) - x^2\phi(x),
\]
so
\[
x^2\phi(x) = \phi(x) - \frac{\rmd}{\rmd x}\bigl(x\phi(x)\bigr),
\]
and hence
\[
\int_0^a x^2\phi(x)\rmd x
= \int_0^a \phi(x)\rmd x - \bigl[x\phi(x)\bigr]_{0}^{a}
= \bigl(\Phi(a) - \tfrac12\bigr) - a\phi(a).
\]
By substituting back into $J(a)$, we have
\begin{align*}
J(a)
&= \frac{a^2}{2}\Phi(-a)
 + \frac{1}{2}\Bigl(\Phi(a) - \frac12 - a\phi(a)\Bigr)\\
&= \frac{a^2}{2}\Phi(-a)
 + \frac{1}{2}\bigl(1 - \Phi(-a)\bigr)
 - \frac{1}{4}
 - \frac{1}{2}a\phi(a)\\
&= \frac{1}{2}(a^2 - 1)\Phi(-a)
 - \frac{1}{2}a\phi(a)
 + \frac{1}{4},
\end{align*}
where we used $\Phi(a) = 1 - \Phi(-a)$ in the second line.

Thus, for any $a>0$,
\[
\int_0^a x\Phi(-x)\rmd x
=
\frac{1}{2}(a^2 - 1)\Phi(-a)
 - \frac{1}{2}a\phi(a)
 + \frac{1}{4}.
\]

Setting $a = h/C$ we obtain
\[
I_T
=
\frac{C^2}{T}
J\p{\frac{h}{C}}
=
\frac{C^2}{T}
\p{
\frac{1}{2}\bigp{\p{h/C}^2 - 1}
      \Phi\p{-\frac{h}{C}}
 - \frac{1}{2}\bigp{h / C}\phi\p{\frac{h}{C}}
 + \frac{1}{4}
},
\]
where $C = \sigma_1(\mu_0) + \sigma_0(\mu_0)$, and $\phi$ and $\Phi$ are the standard normal density and CDF, respectively. 

For large $h$, we have
\[I_T = \frac{C^2}{4T}.\]
Therefore, we have
\begin{align*}
    \liminf_{T\to\infty}T\int_{\mu_0 \in \calM}\int^{\mu_0 + h/\sqrt{T}}_{\mu_0} \big|\mu_1 - \mu_0\big|\Phi\bigp{-U(\bmmu)} \rmd H(\bmmu) \geq \frac{1}{4}\int_{\mu_0 \in \calM}h_1(\mu_0 \mid \mu_0)\bigp{\sigma_1(\mu_0) + \sigma_0(\mu_0)}^2\rmd H_0(\mu_0).
\end{align*}
A similar result holds for the case with $d^\dagger = 0$ as
\begin{align*}
    \liminf_{T\to\infty}T\int_{\mu_1 \in \calM}\int^{\mu_1 + h/\sqrt{T}}_{\mu_1}\big|\mu_1 - \mu_0\big|\Phi\bigp{-U(\bmmu)} \rmd H(\bmmu) \geq \frac{1}{4}\int_{\mu_1 \in \calM}h_1(\mu_1 \mid \mu_1)\bigp{\sigma_1(\mu_1) + \sigma_0(\mu_1)}^2\rmd H_1(\mu_1).
\end{align*}

Finally, we have
\begin{align*}
    \liminf_{T\to\infty}T\int_{\bmmu \in \calM^2} \text{Regret}^\delta_{\bmmu} \rmd H(\bmmu) \geq \frac{1}{4}\sum_{d\in\{1, 0\}}\int_{\mu \in \calM}h_d(\mu \mid \mu_{\backslash d})\bigp{\sigma_1(\mu) + \sigma_0(\mu)}^2\rmd H_{\backslash d}(\mu).
\end{align*}

This completes the proof. 
\end{proof}

\section{Preliminary for the proofs of upper bounds}
In this section, we present preliminary tools for the proofs of our upper bounds. 

\subsection{Almost sure convergence of the first-stage estimator in the treatment allocation phase.}
As a result of the uniform allocation in the treatment allocation phase, the following result holds.

\begin{lemma}
\label{lem:almost_sure_nuisance}
    For any $P_0\in \calP$ and all $d\in\{1, 0\}$, $\widehat{\mu}_{d, rT}\xrightarrow{\mathrm{a.s}} \mu_{a}$ and $\widehat{\sigma}^2_{d, rT} \xrightarrow{\mathrm{a.s}} \sigma^2_d$ as $T \to \infty$.
\end{lemma}
Furthermore, from $\widehat{\sigma}^2_{d, rT} \xrightarrow{\mathrm{a.s}} \sigma^2_d$ and the continuous mapping theorem, $\widehat{w}_{rT}\xrightarrow{\mathrm{a.s}} w^*$ holds. 

\subsection{Upper bound of the probability of misidentification}
We establish an upper bound on the probability of misidentification, $\bbP_{\bmmu}\Bigp{\widehat{\mu}_{d^*_{\bmmu}, T} < \widehat{\mu}_{\widehat{d}^{\delta^{\text{TSNA}}}_T, T}}$. For simplicity, let $\ceil{rT / 2}$ be an integer.

\begin{lemma}
\label{lem:upper_probabilityofmisidentiifcation}
Assume the same conditions as in Proposition~\ref{prp:hahnclt}. 
Let $v_d >0$ be the variance proxy of sub-Gaussian $Y_{d,t}$; that is, it holds that
\[
    \bbE_{\bmmu}\sqb{\exp\bigp{\lambda (Y_{d,t}-\mu_d)}}
    \le \exp\p{\frac{v_d \lambda^2}{2}}
    \quad \forall \lambda \in \bbR.
\]
Then, for $v \coloneqq \max_{d \in \{1, 0\}} v_d$, it holds that
\[
\bbP_{\bmmu}\Bigp{\widehat{\mu}_{d^*_{\bmmu}, T} < \widehat{\mu}_{\widehat{d}^{\delta^{\text{TSNA}}}, T}} \leq 2\exp\p{-\frac{rT \Delta^2_{\bmmu}}{16 v}}.
\]
\end{lemma}

\begin{proof}
For simplicity, without loss of generality, let $d^*_{\bmmu} = 1$ and $\widehat{d}^{\delta^{\text{TSNA}}}_T = 0$. 
We first show that, conditional on the assignment path $D_{1:T} \coloneqq (D_1,\dots,D_T)$, each empirical mean $\widehat{\mu}_{d, T}$ is sub-Gaussian with variance proxy $v / N_d$.

Fix $d\in\{0,1\}$ and condition on $D_{1:T}$. Then
\[
\widehat{\mu}_{d, T} - \mu_d
= \frac{1}{N_d} \sum_{t=1}^T 1\bigsqb{D_t=d}\bigp{Y_{d,t}-\mu_d}.
\]
By the i.i.d.\ structure of $\{(Y_{1,t},Y_{0,t})\}_{t=1}^T$, conditional on $D_{1:T}$ the random variables $\{Y_{d,t} : D_t=d\}$ are i.i.d. with mean $\mu_d$ and the same sub-Gaussian parameter $v$ as the original sequence.

Therefore, for any $\lambda\in\bbR$,
\begin{align*}
\bbE_{\bmmu}\Bigsqb{\exp\bigl(\lambda (\widehat{\mu}_{d, T}-\mu_d)\bigr)\Bigm|D_{1:T}}
&=
\bbE_{\bmmu}\sqb{
\exp\p{\frac{\lambda}{N_d} \sum_{t=1}^T 1[D_t=d](Y_{d,t}-\mu_d)}
\Bigm|D_{1:T}} \\
&=
\prod_{t : D_t=d}
\bbE_{\bmmu}\sqb{
\exp\p{\frac{\lambda}{N_d}(Y_{d,t}-\mu_d)}
\Bigm|D_{1:T}} \\
&\leq
\prod_{t : D_t=d}
\exp\p{\frac{v\lambda^2}{2N_d^2}}
=
\exp\p{\frac{v\lambda^2}{2N_d}}.
\end{align*}
Thus, conditional on $D_{1:T}$, the centered empirical mean $\widehat{\mu}_{d, T}-\mu_d$ is sub-Gaussian with variance proxy $v/N_d$. By the standard Chernoff bound for sub-Gaussian random variables, for any $\varepsilon>0$,
\begin{align}
\label{eq:sg-tail}
\mathbb{P}\p{\widehat{\mu}_{d, T}-\mu_d \leq -\varepsilon \bigm| D_{1:T}}
\le \exp\p{-\frac{N_d \varepsilon^2}{2 v}},
\end{align}
and similarly
\begin{align}
\label{eq:sg-tail-upper}
\mathbb{P}\Bigp{\widehat{\mu}_{d, T}-\mu_d \geq \varepsilon \bigm| D_{1:T}}
\le \exp\p{-\frac{N_d \varepsilon^2}{2 v}}.
\end{align}

Observe that the event $\{\widehat{\mu}_1 < \widehat{\mu}_0\}$ implies that at least one of the two empirical means deviates from its expectation by at least $\Delta/2$ in the wrong direction, where recall that $\Delta = \mu_1 - \mu_0 > 0$. More precisely, it holds that
\begin{align}
\label{eq:event-decomp}
\Bigcb{\widehat{\mu}_{1, T} < \widehat{\mu}_{0, T}}
\subset
\cb{\widehat{\mu}_{1, T} \le \mu_1 - \frac{\Delta}{2}}
\cup
\cb{\widehat{\mu}_{0, T} \ge \mu_0 + \frac{\Delta}{2}}.
\end{align}
Indeed, if both
\[
\widehat{\mu}_{1, T} > \mu_1 - \frac{\Delta}{2}
\quad\text{and}\quad
\widehat{\mu}_{1, T} < \mu_0 + \frac{\Delta}{2}
\]
hold, then
\[
\widehat{\mu}_{1, T} - \widehat{\mu}_{0, T}
>
\mu_1 - \frac{\Delta}{2}
-
\p{\mu_0 + \frac{\Delta}{2}}
=
\mu_1 - \mu_0 - \Delta
= 0,
\]
which contradicts $\widehat{\mu}_{1, T} < \widehat{\mu}_{0, T}$.

Combining \eqref{eq:event-decomp} with the union bound and the sub-Gaussian tail bounds \eqref{eq:sg-tail} and \eqref{eq:sg-tail-upper} with $\varepsilon = \Delta/2$ gives, conditional on $D_{1:T}$, we have
\begin{align*}
\mathbb{P}\Bigp{\widehat{\mu}_1 < \widehat{\mu}_0 \bigm| D_{1:T}}
&\le
\mathbb{P}\Bigp{\widehat{\mu}_1 \le \mu_1 - \frac{\Delta}{2} \Bigm| D_{1:T}}
+
\mathbb{P}\Bigp{\widehat{\mu}_0 \ge \mu_0 + \frac{\Delta}{2} \Bigm| D_{1:T}} \\
&\le
\exp\p{-\frac{N_1 \Delta^2_{\bmmu}}{8 v}}
+
\exp\p{-\frac{N_0 \Delta^2_{\bmmu}}{8 v_0}},
\end{align*}
where $N_d$ is defined as 
\[N_d \coloneqq \sum^T_{t=1}\mathbbm{1}[D_t = d].\]

Taking expectations with respect to $D_{1:T}$ yields
\begin{align}
\label{eq:before-lower-bound}
\mathbb{P}\Bigp{\widehat{\mu}_{1, T} < \widehat{\mu}_{0, T}}
&=
\bbE\sqb{\mathbb{P}\Bigp{\widehat{\mu}_1 < \widehat{\mu}_0 \bigm| D_{1:T}}}\nonumber\\
&\le
\bbE\Bigsqb{
\exp\left(-\frac{N_1 \Delta^2_{\bmmu}}{8 v}\right)} + \bbE\Bigsqb{
\exp\left(-\frac{N_0 \Delta^2_{\bmmu}}{8 v}\right)}.
\end{align}

Since $N_d \ge rT/2$, for each $d\in\{1, 0\}$ we have
\[
\exp\left(-\frac{N_1 \Delta^2_{\bmmu}}{8 v}\right)
\le
\exp\left(-\frac{rT / 2 \Delta^2_{\bmmu}}{8 v}\right)
=
\exp\left(-\frac{rT \Delta^2_{\bmmu}}{16 v}\right)\]
Substituting this into \eqref{eq:before-lower-bound}, and dropping the expectation, we obtain
\[
\mathbb{P}\Bigp{\widehat{\mu}_{1, T} < \widehat{\mu}_{0, T}}
\le
2\exp\left(-\frac{rT \Delta^2_{\bmmu}}{16 v}\right).\]
This completes the proof. 
\end{proof}

\section{Proof of the worst-case upper bound (Theorem~\ref{thm:minimax_upperbound})}
\label{appdx:proof:worstcase_upper}

In this section, we prove Theorem~\ref{thm:minimax_upperbound}.
The key ingredients are the almost sure convergence of the empirical allocation (Lemma~\ref{lem:almost_sure_nuisance}) and the exponential upper bound on the misidentification probability (Lemma~\ref{lem:upper_probabilityofmisidentiifcation}).

\begin{proof}
Fix $\bmmu \in \calM^2$ and, without loss of generality, assume that $d^*_{\bmmu} = 1$.
Then the simple regret of $\delta^{\text{TSNA}}$ can be written as
\[
  \text{Regret}^{\delta^{\text{TSNA}}}_{\bmmu}
  = \Delta_{\bmmu}\bbP_{\bmmu}\left(\widehat{d}^{\delta^{\text{TSNA}}}_T = 0\right)
  = (\mu_1 - \mu_0)\bbP_{\bmmu}\left(\widehat{\mu}_{1,T} \le \widehat{\mu}_{0,T}\right).
\]

Under the TSNA experiment, Lemma~\ref{lem:almost_sure_nuisance} implies that the empirical allocation converges almost surely to some allocation ratio $w$, that is,
\[
\frac{1}{T}\sum_{t=1}^T \mathbbm{1}[D_t = 1] \xrightarrow{\mathrm{a.s.}} w.
\]
Moreover, under the condition $r / 2 \le \min_{d\in\{1, 0\}} \sigma_d(\mu)/\big(\sigma_1(\mu)+\sigma_0(\mu)\big)$ in Theorem~\ref{thm:minimax_upperbound}, the second stage corrects for the first-stage uniform allocation, and the limiting allocation coincides with the ideal ratio
\[
w^* = \frac{\sigma_1(\mu)}{\sigma_1(\mu) + \sigma_0(\mu)}.
\]
Hence, Proposition~\ref{prp:hahnclt} and Lemma~\ref{lem:upper_probabilityofmisidentiifcation} applies with these weights.

From Lemma~\ref{lem:upper_probabilityofmisidentiifcation}, there exists $v > 0$ independent of $T$ such that 
\[
\bbP_{\bmmu}\Bigp{\widehat{\mu}_{1,T} \le \widehat{\mu}_{0,T}}
\leq 
2\exp\p{-\frac{rT \Delta^2_{\bmmu}}{16 v}}.
\]

Therefore, we have
\begin{align*}
\sqrt{T}\text{Regret}^{\delta^{\text{TSNA}}}_{\bmmu}
&= \sqrt{T}(\mu_1 - \mu_0)
\bbP_{\bmmu}\left(\widehat{\mu}_{1,T} \le \widehat{\mu}_{0,T}\right)\\
&\le 2\sqrt{T}(\mu_1 - \mu_0)
 \exp\p{-\frac{rT \Delta^2_{\bmmu}}{16 v}}
 + o(1).
\end{align*}
The right-hand side is of the form
\[
f_T(\Delta) \coloneqq \sqrt{T}\Delta
\exp\p{-\frac{rT \Delta^2_{\bmmu}}{16 v}},
\qquad \Delta = \mu_1 - \mu_0.
\]
When $\sqrt{T}\Delta \to \infty$, then $f_T(\Delta)$ converges to zero with an exponential rate. Therefore, we only consider the case with $\Delta = O(1/\sqrt{T})$. 

For $\mu_1 - \mu_0 = h/\sqrt{T}$, from Proposition~\ref{prp:hahnclt}, we have
\begin{align*}
\sqrt{T}\text{Regret}^{\delta^{\text{TSNA}}}_{\bmmu}
&= \sqrt{T}(\mu_1 - \mu_0)
\bbP_{\bmmu}\left(\widehat{\mu}_{1,T} \le \widehat{\mu}_{0,T}\right)\\
&\le h\Phi\p{ - \frac{h}{\sqrt{V\p{w^*}}}} + \epsilon.
\end{align*}
The right-hand side is of the form
\[
g_T(h) \coloneqq h
\Phi\p{ - \frac{h}{\sqrt{V\p{w^*}}}}.
\]
The maximizer of $g_T(h)$
over $h > 0$ is given as
\[h = \sqrt{V(w^*)}.\]
Therefore, we have
\begin{align*}
    \limsup_{T\to\infty}\sup_{\bmmu\in\calM^2}\sqrt{T}\text{Regret}^{\delta^{\text{TSNA}}}_{\bmmu} \leq \sqrt{V\p{w^*}}\Phi\p{-\frac{\sqrt{V\p{w^*}}}{\sqrt{V\p{w^*}}}} = \sqrt{V\p{w^*}}\Phi(- 1).
\end{align*}
This completes the proof. 
\end{proof}

\section{Proof of the average upper bound (Theorem~\ref{thm:bayes_upperbound})}
\label{appdx:proof:average_upper}
We prove the average upper bound.

\begin{proof}
The regret can be decomposed as 
\begin{align*}
    &\text{Regret}^{\delta^{\text{TSNA}}}_{\bmmu} = \bbE_{\bmmu}\sqb{\mu_{d^*_{\bmmu}} - \mu_{\widehat{d}_T^{\delta}}} = \Delta_{\bmmu}\bbP_{\bmmu}\p{\widehat{d}_T^{\delta} \neq d^*_{\bmmu}}. 
\end{align*}
We consider bounding
\begin{align}
    \limsup_{T\to \infty}T \int_{\bmmu\in \calM^2} \Delta_{\bmmu}\bbP_{\bmmu}\p{\widehat{d}_T^{\delta} \neq d^*_{\bmmu}} \rmd H\bigp{\bmmu}.
\end{align}

We first bound
\[
T \int_{\bmmu\in \calM^2} \mathbbm{1}\bigsqb{\mu_1 \geq \mu_0}\Delta_{\bmmu}\bbP_{\bmmu}\p{\widehat{d}_T^{\delta} \neq d^*_{\bmmu}}\rmd H\bigp{\bmmu}.
\]
We have
\begin{align*}
    &T \int_{\bmmu\in \calM^2} \mathbbm{1}\bigsqb{\mu_1 \geq \mu_0}\Delta_{\bmmu}\bbP_{\bmmu}\p{\widehat{d}_T^{\delta} \neq d^*_{\bmmu}}\rmd H\bigp{\bmmu}\\
    &= T \int_{\mu_0\in \calM} \int_{\mu_1\in \calM} \mathbbm{1}\bigsqb{\mu_1 \geq \mu_0}\Delta_{\bmmu}\bbP_{\bmmu}\p{\widehat{d}_T^{\delta} \neq d^*_{\bmmu}}\rmd H\bigp{\bmmu}\\
    &= T \int_{\mu_0\in \calM} \int_{\mu_1\in \calM} \mathbbm{1}\bigsqb{\mu_1 \geq \mu_0}\Delta_{\bmmu}\bbP_{\bmmu}\p{\widehat{d}_T^{\delta} \neq d^*_{\bmmu}}h_1(\mu_1 \mid \mu_0) \rmd \mu_1 \rmd H_0\bigp{\mu_0}\\
    &= (1 + o(1))T \int_{\mu_0\in \calM} h_1(\mu_0 \mid \mu_0) \int_{\mu_1\in \calM} \mathbbm{1}\bigsqb{\mu_1 \geq \mu_0}\Delta_{\bmmu}\bbP_{\bmmu}\bigp{\widehat{\mu}_{1, T} \leq \widehat{\mu}_{0, T}} \rmd \mu_1 \rmd H_0\bigp{\mu_0}\\
    &= (1 + o(1))T \int_{\mu_0\in \calM} h_1(\mu_0 \mid \mu_0) \int_{\mu_1\in \calM} \mathbbm{1}\bigsqb{\mu_1 \geq \mu_0}\Delta_{\bmmu}\bbP_{\bmmu}\bigp{\widehat{\mu}_{1, T} \leq \widehat{\mu}_{0, T}} \rmd \mu_1 \rmd H_0\bigp{\mu_0}.
\end{align*}

Fix $h > 0$ such that $\mu_0 + h / \sqrt{T} \in \calM$. We consider bounding
\[
\int_{\mu_0 \in \calM}\Delta_{\bmmu}\bbP_{\bmmu}\bigp{\widehat{\mu}_{1, T} \leq \widehat{\mu}_{0, T}} \rmd \mu_1.
\]
From Proposition~\ref{prp:hahnclt} and Lemma~\ref{lem:upper_probabilityofmisidentiifcation}, there exists $v > 0$ independent of $T$ such that 
\begin{align*}
    &\int_{\mu_0 \in \calM}\Delta_{\bmmu}\bbP_{\bmmu}\bigp{\widehat{\mu}_{1, T} \leq \widehat{\mu}_{0, T}} \rmd \mu_1\\
    &\leq \int^{\mu_0 + h / \sqrt{T}}_{\mu_0}\bigp{\mu_1 - \mu_0}\p{\Phi\p{ - \frac{\sqrt{T}\bigp{\mu_1 - \mu_0}}{\sqrt{V\p{w^*}}}} + \epsilon}\rmd \mu_1\\
    &\ \ \ \ \ \ + 2\int_{\mu_1 > \mu_0 + h / \sqrt{T}}\bigp{\mu_1 - \mu_0}\exp\p{-\frac{rT \bigp{\mu_1 - \mu_0}^2}{16 v}}\rmd \mu_1.
\end{align*}

Let us define
\begin{align*}
    &I_1(h) \coloneqq \int_{\mu_0}^{\mu_0 + h / \sqrt{T}}\bigp{\mu_1 - \mu_0}\p{\Phi\p{ - \frac{\sqrt{T}\bigp{\mu_1 - \mu_0}}{\sqrt{V_1\p{w^*}}}} + \epsilon}\rmd \mu_1,\\
    &I_2(h) \coloneqq \int_{\mu_1 > \mu_0 + h / \sqrt{T}}\bigp{\mu_1 - \mu_0}\exp\p{-\frac{rT \bigp{\mu_1 - \mu_0}^2}{16 v}}\rmd \mu_1,
\end{align*}
where 
\[V_1\p{w^*} \coloneqq \bigp{\sigma_1(\mu_0) + \sigma_0(\mu_0)}.\]

Here we set $\Delta = \mu_1 - \mu_0$, so that
\begin{align*}
    I_1(h)
    &= \int_0^{h/\sqrt{T}} \Delta \p{\Phi\p{-\frac{\sqrt{T}\Delta}{\sqrt{V_1\p{w^*}}}} + \epsilon}\rmd\Delta\\
    &= \int_0^{h/\sqrt{T}} \Delta \Phi\p{-\frac{\sqrt{T}\Delta}{\sqrt{V_1\p{w^*}}}}\rmd\Delta
       + \epsilon \int_0^{h/\sqrt{T}} \Delta \rmd\Delta.
\end{align*}

First, we focus on the main term that ignores the $\epsilon$ part and define
\[
    J(h) \coloneqq \int_0^{h/\sqrt{T}} \Delta \Phi\p{-\frac{\sqrt{T}\Delta}{\sqrt{V_1\p{w^*}}}}\rmd\Delta.
\]
Let $a \coloneqq \sqrt{T}/\sqrt{V\p{w^*}}$ and change variables by $x = a\Delta$. Then
\begin{align*}
    J(h)
    &= \frac{1}{a^2}\int_0^{ah/\sqrt{T}} x \Phi(-x)\rmd x
    = \frac{V\p{w^*}}{T}\int_0^{h/\sqrt{V_1\p{w^*}}} x \Phi(-x)\rmd x.
\end{align*}
If we extend the upper limit of the integral to infinity, we obtain
\begin{align*}
    \int_0^{\infty} x \Phi(-x)\rmd x
    &= \sqb{\frac{x^2}{2}\Phi(-x)}_{0}^{\infty}
       + \frac12 \int_0^{\infty} x^2 \varphi(x)\rmd x\\
    &= 0 + \frac{1}{2} \cdot \frac{1}{2} = \frac{1}{4},
\end{align*}
where $\varphi(x)$ is the standard normal density and we use
$\int_{-\infty}^{\infty} x^2 \varphi(x)\rmd x = 1$ and symmetry.
Therefore,
\begin{align*}
    \limsup_{h\to\infty} J(h)
    &= \frac{V_1\p{w^*}}{T} \int_0^{\infty} x \Phi(-x)\rmd x
     = \frac{V_1\p{w^*}}{T} \cdot \frac14
     = \frac{V_1\p{w^*}}{4T}.
\end{align*}
Next, for $I_2(h)$, let $C \coloneqq \tfrac{r}{16v} > 0$. Then
\begin{align*}
    I_2(h)
    &= \int_{h/\sqrt{T}}^{\infty} \delta
       \exp\p{-TC\Delta^2_{\bmmu}}\rmd\delta\\
    &= \frac{1}{2TC}\exp\bigl(-C h^2\bigr),
\end{align*}
so $I_2(h) \to 0$ as $h \to \infty$.

Combining these displays and neglecting the $\epsilon$ term, for large $T$, we obtain
\begin{align*}
    \int_{\mu_1 \in \calM}\Delta_{\bmmu}\bbP_{\bmmu}\bigp{\widehat{\mu}_{1, T} \leq \widehat{\mu}_{0, T}} \rmd \mu_1
    \leq
    \frac{V_1\p{w^*}}{4T}.
\end{align*}
Similarly, for large $T$, we can show that
\begin{align*}
    \int_{\mu_0 \in \calM}\Delta_{\bmmu}\bbP_{\bmmu}\bigp{\widehat{\mu}_{1, T} \geq \widehat{\mu}_{0, T}} \rmd \mu_0
    \leq 
    \frac{V_0\p{w^*}}{4T},
\end{align*}
where
\[V_0\p{w^*} \coloneqq \bigp{\sigma_1(\mu_1) + \sigma_0(\mu_1)}.\]

Combining these upper bounds, we have
\begin{align*}
    &\limsup_{T\to \infty}T \int_{\bmmu\in \calM^2} \text{Regret}^{\delta^{\text{TSNA}}}_{\bmmu} \rmd H\bigp{\bmmu}\\
    &\leq (1 + o(1))\frac{1}{4}\sum_{d\in\{1, 0\}}\int_{\mu \in \calM}h_d(\mu \mid \mu_{\backslash d})\bigp{\sigma_1(\mu) + \sigma_0(\mu)}^2\rmd H_{\backslash d}(\mu).
\end{align*}
This completes the proof. 
\end{proof}

\end{document}